\documentclass[sigconf, authorversion]{acmart}

\usepackage{graphicx}

\usepackage{xcolor}
\usepackage{amsmath}
\usepackage{multirow}
\usepackage{longtable}
\usepackage[font=footnotesize]{subcaption}
\usepackage{multicol}
\usepackage{longtable}
\usepackage{lipsum}
\usepackage{balance}

\AtBeginDocument{%
  \providecommand\BibTeX{{%
    \normalfont B\kern-0.5em{\scshape i\kern-0.25em b}\kern-0.8em\TeX}}}

\copyrightyear{2020} 
\acmYear{2020} 
\setcopyright{acmcopyright}\acmConference[WiSec '20]{13th ACM Conference on Security and Privacy in Wireless and Mobile Networks}{July 8--10, 2020}{Linz (Virtual Event), Austria}
\acmBooktitle{13th ACM Conference on Security and Privacy in Wireless and Mobile Networks (WiSec '20), July 8--10, 2020, Linz (Virtual Event), Austria}
\acmPrice{15.00}
\acmDOI{10.1145/3395351.3399357}
\acmISBN{978-1-4503-8006-5/20/07}

\begin{document}

\title{Fingerprinting Encrypted Voice Traffic on Smart Speakers with Deep Learning}


\author{Chenggang Wang}
\authornote{The first two authors contributed equally to this research.}
\affiliation{%
  \institution{University of Cincinnati, USA}
}
\email{wang2c9@mail.uc.edu}

\author{Sean Kennedy}
\affiliation{%
  \institution{University of Cincinnati, USA}
}
\email{kenneds6@mail.uc.edu}

\author{Haipeng Li}
\affiliation{%
  \institution{University of Cincinnati, USA}
}
\email{li2hp@mail.uc.edu}

\author{King Hudson}
\affiliation{%
  \institution{University of Cincinnati, USA}
}
\email{hudsonk4@mail.uc.edu}

\author{Gowtham Atluri}
\affiliation{%
  \institution{University of Cincinnati, USA}
}
\email{atlurigm@ucmail.uc.edu}

\author{Xuetao Wei}
\affiliation{%
  \institution{Southern University of Science and Technology, China}
}
\email{weixt@sustech.edu.cn}

\author{Wenhai Sun}
\affiliation{%
  \institution{Purdue University, USA}
}
\email{whsun@purdue.edu}

\author{Boyang Wang}
\affiliation{%
  \institution{University of Cincinnati, USA}
}
\email{boyang.wang@uc.edu}

\renewcommand{\shortauthors}{C. Wang, S. Kennedy, H. Li, K. Hudson, G. Atluri, X. Wei, W. Sun and B. Wang}

\begin{abstract}
This paper investigates the privacy leakage of smart speakers under an encrypted traffic analysis attack, referred to as voice command fingerprinting. In this attack, an adversary can eavesdrop both outgoing and incoming encrypted voice traffic of a smart speaker, and infers which voice command a user says over encrypted traffic. 
We first built an automatic voice traffic collection tool and collected two large-scale datasets on two smart speakers, Amazon Echo and Google Home. 
Then, we implemented proof-of-concept attacks by leveraging deep learning. Our experimental results over the two datasets indicate disturbing privacy concerns. Specifically, compared to 1\% accuracy with random guess, our attacks can correctly infer voice commands over encrypted traffic with 92.89\% accuracy on Amazon Echo. 

Despite variances that human voices may cause on outgoing traffic, our proof-of-concept attacks remain effective even only leveraging incoming traffic (i.e., the traffic from the server). This is because the AI-based voice services running on the server side response commands in the same voice and with a deterministic or predictable manner in text, which leave distinguishable pattern over encrypted traffic. 
We also built a proof-of-concept defense to obfuscate encrypted traffic. Our results show that the defense can effectively mitigate attack accuracy on Amazon Echo to 
32.18\%. 
\end{abstract}



\maketitle


\section{Introduction}
\label{sec:intro}

Smart speakers, such as Amazon Echo, Google Home and Apple HomePod, are being increasingly adopted in the U.S. with sales surpassing 133 million \cite{SmartMarket}. However, privacy remains one of the major concerns limiting a more widespread adoption among consumers. This includes two types of concerns: (1) privacy disclosed to voice service providers \cite{AmazonStaff}, and (2) the focus of this research, sensitive information that can be revealed by external attackers.  

We investigated the privacy leakage of smart speakers by considering an external attacker that runs voice command fingerprinting attacks \cite{KLWLWS19}. In this attack, an attacker eavesdrops encrypted voice traffic of a smart speaker, and leverages side-channel information, including the size, direction, and order of encrypted packets, to infer a user's voice command without decryption. For instance, an attacker can be a \textit{local eavesdropper} on a victim's WiFi network or a compromised WiFi access point. 
As the content of a response from the server is correlated with a voice command, an attacker leverages both outgoing traffic (the encrypted packets of a \textit{voice command}) and incoming traffic (the encrypted packets of a \textit{response}) to infer a voice command in this attack.  

Revealing voice commands can uncover users' activities, lead to unauthorized 
disclosure, and compromise privacy of millions of users. 
Moreover, an attacker could leverage voice command fingerprinting to assist malicious attacks (such as \textit{skill squatting} \cite{KPMHMBB18, ZMFWTQ19}) to attack specific targets. For instance, an attacker could infer which voice commands a specific victim often says by leveraging voice command fingerprinting and then create malicious skills using skill squatting, where the names of malicious skills share similar pronunciations of words appeared in those voice commands. An attacker could further record user conversations through smart speakers using malicious skills and steal sensitive information such as passwords and credit card information \cite{KPMHMBB18, ZMFWTQ19}.     

Similar to website fingerprinting \cite{LL06, HWF09, PNZE11, DCRS12, WCNJG14, AG16, HD16, PLZHPWE16, SIJW18, RPJGJ18, OSH18, SKHMMOY19, KRHP19, SMRW19, BLKD19} 
voice command fingerprinting is an encrypted traffic analysis attack, which can be formulated as supervised learning problem. Machine-learning-based encrypted traffic analysis can also be used to fingerprint devices \cite{URCBB13, AHRNF19}. Kennedy et al. \cite{KLWLWS19} previously studied voice command fingerprinting over a small dataset, which consists of 100 commands with 10 traffic traces per command. By manually selecting features and utilizing AdaBoost as the classifier, their attack 
can achieve 33.8\% accuracy. 

\begin{figure*}[ht]
\centering 
\includegraphics[width=14cm]{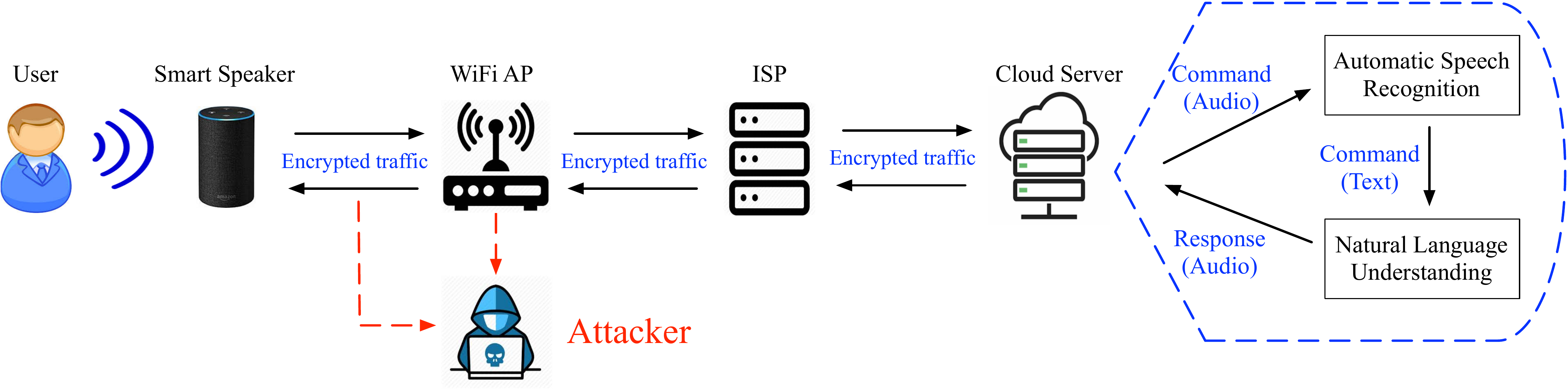} 
\vspace{-5pt}
\caption{The system model of voice command fingerprinting attacks. }
\vspace{-5pt}
\label{fig:model}
\end{figure*}

\textit{This paper aims to improve attack accuracy and advance our understanding of the privacy leakage.} Deep learning techniques \cite{LBH15} that have been found to improve attack accuracy of encrypted traffic analysis in website fingerprinting \cite{AG16, SIJW18, RPJGJ18, OSH18, SKHMMOY19}, 
are potential candidates to address the problem of voice command fingerprinting. However, deep learning techniques require large datasets with thousands to millions of samples. Currently, there are no large-scale research datasets available that can be harnessed in the context of voice command fingerprinting. 

In this paper, we report the advances we made in bridging the gap in the understanding of privacy impacts of smart speakers. Our experimental results derived from neural networks over two large-scale datasets indicate disturbing privacy concerns on smart speakers as well as the driving force of smart speakers --- the AI-based voice services. According to our results, despite high variances that human voices may cause on the outgoing traffic of a smart speaker, \textbf{our proof-of-concept attacks remain effective by only leveraging the incoming traffic from the server side. }  \textit{This is mainly because the AI-based voice services running on the server side response commands in the same voice and with a deterministic or predictable manner in text, which leave distinguishable network traffic pattern.}

\textbf{Contributions.} Our main contributions are summarized below:
\begin{itemize}
\item We built an automatic tool to collect encrypted voice traffic on a smart speaker. 
It is capable of collecting approximately 3,000 traffic traces per day, which addresses the limitation of data collection in voice command fingerprinting. 
\item We collected two large-scale datasets on two popular smart speakers, Amazon Echo and Google Home, using 5 automated voices rendered by public text-to-speech APIs. 
Each dataset consists of 150,000 traffic traces including 100 commands/classes and 1,500 traffic traces per class. 

\item We performed the attack utilizing Convolutional Neural Networks (CNN), Long Short-Term Memory (LSTM) and Stacked Autoencoder (SAE) respectively. Our results using bidirectional traffic of Amazon Echo in \textit{the closed-world setting}\footnote{In the closed world setting, an attacker knows a traffic is associated with a given list of commands and infers which command it is. In the open-world setting, an attacker decides whether a traffic is associated with a given list of commands.} show that, 
CNN attained \textbf{89.05\%} accuracy, LSTM achieved 88.65\% and SAE obtained 75.98\%. Both CNN and LSTM outperformed the previous attack in \cite{KLWLWS19}. 
Our attack using ensemble learning \cite{OM99} further improved accuracy to \textbf{92.89\%}. %
\item As human voices vary in practice due to age, gender and accent, which could cause higher variances on outgoing traffic than the automated voices we utilized in our data collection. We also demonstrated that our attacks are effective using incoming traffic only. Specifically, both CNN and LSTM still achieved over \textbf{81\%} accuracy in the closed world-setting on Amazon Echo dataset. Our attack results in  \textit{the open-world setting} are also highly effective.

\item We designed a proof-of-concept defense, which obfuscates traffic and mitigates privacy leakage against voice command fingerprinting. 
According to our results on Amazon Echo dataset, with privacy parameter $\epsilon=0.005$, our defense can reduce attack accuracy to \textbf{1.23\%} if an attacker trains models with original traffic and tests with obfuscated traffic. If an attacker adapts, in which it trains and tests with obfuscated traffic, attack accuracy can still be reduced to  \textbf{32.18\%}. 
\end{itemize}


\section{Related Work}
\label{sec:related}

\textbf{Website Fingerprinting.} The purpose of website fingerprinting is to infer which website a user visits over encrypted traffic \cite{LL06, HWF09, PNZE11, DCRS12, WCNJG14, HD16, PLZHPWE16, SIJW18, RPJGJ18}. Early-stage research in website fingerprinting focused on manually extracting features from encrypted traffic and harnessing different conventional machine learning algorithms to achieve higher accuracy. Among these studies, Panchenko et al. \cite{PLZHPWE16} proposed CUMUL by considering the cumulative sum function of traffic size. This attack leveraged Support Vector Machine as the classifier and outperformed other methods \cite{RPJGJ18}. This attack achieved an accuracy of 93\%, and is even comparable in performance with deep-learning-based attacks \cite{SIJW18, RPJGJ18}. 

Recent work in website fingerprinting attacks used deep learning models to automatically extract features and resulted in higher accuracy. Sirinam et al. \cite{SIJW18} leveraged Convolutional Neural Network (CNN) and attained 98\% accuracy in the closed-world setting with 95 websites. 
Rimmer et al. \cite{RPJGJ18} investigated website fingerprinting with CNN, Long Short-Term Memory (LSTM) and Stacked Denoising Autoencoder (SDAE). SDAE achieved 94\% accuracy on a dataset with 900 classes while CNN and LSTM reached 91\% and 88\%. 

Different defense methods have also been proposed. 
BuFLO 
\cite{DCRS12} sends packets at a fixed size with fixed intervals but introduces high latency. 
Juarez et al. \cite{JIPDW16} designed WTF-PAD to obfuscate traffic pattern by using adaptive padding \cite{SW06}, which hides traffic gaps and introduces no latency. Wang et al. devised Walkie-Talkie \cite{WG17}, which applied half-duplex model and burst molding to Tor traffic. 

Both WTF-PAD \cite{JIPDW16} and Walkie-Talkie \cite{WG17} cause low latency, but can be compromised by CNN-based attacks \cite{SIJW18}. A CNN-based attack achieved 90\% accuracy against WTF-PAD and 49.7\% accuracy against Walkie-Talkie. Imani et al. \cite{IRMW19} proposed to leverage adversarial examples \cite{SZSBEGF14} 
as a defense against website fingerprinting. However, it requires the knowledge of entire traffic traces in advance, and hence cannot obfuscate traffic on the fly.  

\textbf{Video Stream Fingerprinting.} Schuster et al. \cite{SST17} showed that traffic bursts are useful to identify encrypted MPEG-DASH video streams. Zhang et al. \cite{ZHRZ19} demonstrated the threat of video stream fingerprinting with 40 YouTube videos with 100 traces per video. CNN achieved 95\% accuracy and outperformed competing approaches. On the other hand, the authors showed that leveraging differential privacy on time-series data (e.g., $d^{*}$-privacy \cite{XRZ15}) can obfuscate traffic pattern and preserve privacy over encrypted traffic. In contrast to website fingerprinting, which extracts features on bi-directional traffic, video stream fingerprinting leverages one-way traffic (i.e., encrypted video traffic sent by the server).  

\textbf{Voice Command Fingerprinting.} Kennedy et al. investigated voice command fingerprinting attacks over a small dataset, which includes 100 classes and 10 traces per class \cite{KLWLWS19}. The authors examined several traditional machine learning methods in website fingerprinting and applied to encrypted voice traffic on smart speakers. A method 
leveraging AdaBoost achieved over 33.8\% accuracy in the closed-world setting and outperformed others. The features that the authors examined in \cite{KLWLWS19} include the size of each burst, total transmitted bytes, the number of bursts, occurring packet sizes, percentage incoming packets, and the number of packets. 

Apthorpe et al. \cite{AHRNF19} inferred user activities at home by identifying different smart home devices (including smart speakers) through encrypted traffic pattern. In their study, the authors inferred 3 voice commands (with 3 samples per command) on Amazon Echo based on traffic rate. The dataset and attack methods in \cite{AHRNF19} are not as comprehensive as our study. 

\textbf{Fingerprinting on Encrypted VoIP Traffic.} Several previous studies examined the privacy of encrypted VoIP (Voice over IP traffic) \cite{WCJ05, WBMM07, WBCMM08, BDDK10, WMSM11}. 
These studies leverage packet size as the only fingerprint because VoIP uses a combination of Variable Bit Rate encoding with stream cipher, 
but are not effective if data is encrypted with block cipher \cite{WBCMM08}. These attacks are not applicable to voice command fingerprinting as voice traffic on smart speakers is encrypted with block cipher.

\textbf{Fingerprinting IoT Devices.} Many research studies \cite{AFASMACSU18, BBPSRR18, JOAWQ18, AHRNF19, MSE19, TDSBG19, AGY20, TVMD20, MS20} have investigated how to identify IoT devices as well as associated events within a smart home by analyzing encrypted traffic. For instance, Acar et. al. proposed a multi-stage attack, which can achieve over 90\% accuracy inferring smart home devices. Different from these studies, our attack focuses on analyzing privacy within a single type of smart home devices.

\textbf{Other Attacks on Smart Speakers.}
Injection attacks can inject voice commands through similar pronunciations \cite{KPMHMBB18, ZMFWTQ19}, audible sounds \cite{ZYJZZX17, RSHC18}, or songs \cite{YCZLLCZHWG18}. These attacks focus on the vulnerabilities of Automatic Speech Recognition. Abdullah et al. \cite{AGPTBW19} explored the vulnerabilities of the signal processing step before Automatic Speech Recognition. Zhang et al. \cite{ZXMYCG19} investigated the vulnerability of Natural Language Understanding. 
In contrast to these active attacks, voice command fingerprinting is passive. 

\section{Background}
\label{sec:background}

\textbf{System Model.} The system model is illustrated in Fig.~\ref{fig:model}. It includes four entities: a smart speaker, a WiFi access point, an Internet Service Provider and a server. When a smart speaker receives a user's voice command upon hearing a wake word, it records the voice data and then forwards data to its cloud server. The voice services on the server side will produce responses to a voice command. 
The voice data traffic between a smart speaker and the server is protected by off-the-shelf encryption technology. For instance, Amazon Echo leverages TLS (Transport Layer Security) 1.2 and all traffic is encrypted with AES (Advanced Standard Encryption) \cite{KLWLWS19}. 


\textbf{Threat Model.} We assume an attacker is a \textit{local eavesdropper} who can sniff the network traffic of a smart speaker. For instance, an attacker can be an eavesdropper on the victim’s WiFi network or a compromised WiFi access point. This attacker cannot decrypt encrypted packets. In addition, this attacker does not drop, change or inject packets.

We assume there is one smart speaker in the victim's WiFi network. We assume the attacker knows the model of a smart speaker, e.g., Amazon Echo or Google Home. We assume the attacker can infer the IP address of a smart speaker as well as the IP address of the server running voice services. With the IP address of a smart speaker, the attacker can filter out traffic from other devices connecting to the same WiFi access point \cite{AHRNF19}. 

\textbf{\textit{Inferring the IP address of a smart speaker.}} Inferring the IP of a smart speaker is feasible for a local eavesdropper. Many existing studies \cite{AFASMACSU18, BBPSRR18, JOAWQ18, AHRNF19, MSE19, TDSBG19, AGY20, TVMD20, MS20} have shown that it is possible to distinguish the traffic of a smart speaker (therefore its IP address) from other smart home devices over encrypted traffic. 

Once the IP address of a smart speaker is inferred, the server's IP address can be easily observed as a smart speaker mainly communicates with its voice services. Moreover, DNS queries from a smart speaker can also be used to infer the IP addresses of its voice services. For instance, Amazon Echo in our data collection sent DNS queries to resolve the IP addresses of domain name \texttt{unagi-na.amazon.com}. As DNS queries and responses are not encrypted, and the IP addresses in the answer section of a DNS response can be easily obtained (e.g., using \texttt{Wireshark} or \texttt{dig} command).  

\textbf{\textit{A traffic trace contains all of the packets for a voice command and its response}}. We assume packets that are sent to the server are outgoing packets (packets containing a voice command), and packets that are sent to a smart speaker are incoming packets (packets containing a response). We assume that an attacker can infer the start time and the end time of each traffic trace\footnote{It is feasible for an attacker to infer the start time of each trace on a smart speaker. For example, through our data collection, there is a significant amount of outgoing traffic initiated around the start time of a traffic trace. The end time could be identified once there is no significant volume of traffic after certain time frame, e.g., 2$\sim$3 seconds.}. An attacker can learn side-channel information, including direction, packet size, and timestamp. A traffic trace of a voice command $C$ and its response can be described as
\begin{equation}
T_{C} = \langle (b_{1}, s_{1}, t_{1}), ..., (b_{n}, s_{n}, t_{n}) \rangle
\label{equation:genetric}
\end{equation}
where $n$ is the number of packets in this trace. Each direction $b_{i}$ is either $+1$ (outgoing) or $-1$ (incoming), each packet size $s_{i}$ is in bytes, and each timestamp $t_{i}$ is represented in milliseconds. 

\textbf{Closed-World Setting and Open-World Setting.} We investigate voice command fingerprinting attacks in both the closed-world setting and the open-world setting.

In the closed-world setting, we assume that an attacker has a prior set of voice commands. For example, this prior set of voice commands can be a set of popular commands that users would ask. Given this prior set, an attacker can harvest labeled traffic traces for each voice command by itself. An attacker can capture an unlabeled traffic trace from a user’s smart speaker. This unlabeled traffic trace is associated with one of the voice commands in the prior set. The objective of this attacker is to infer which voice command this unlabeled trace is associated with. 

In the open-world setting, an unlabeled traffic trace may not be in the prior set. The objective of this attacker is to infer whether the voice command of this unlabeled traffic trace is in the prior set. 

\textbf{Privacy Metric.} We leverage the accuracy of the classification to determine the privacy leakage under voice command fingerprinting in the closed-world setting. For the open-world setting, we leverage true positive rate and false positive rate. This privacy metric is often used in the literature of encrypted traffic analysis.


\section{Voice Command Fingerprinting} 
\label{sec:attack}

\subsection{Automatic Traffic Collection Tool }

Our first challenge is the lack of sufficient training data (or the lack of tools to collect sufficient training data).  
Existing data collection methods can automatically capture web traffic. However, they do not directly apply to the traffic collection on smart speakers, where \textit{voice interactions are required}. 

\begin{figure}[t]
\centering
\includegraphics[width=7cm]{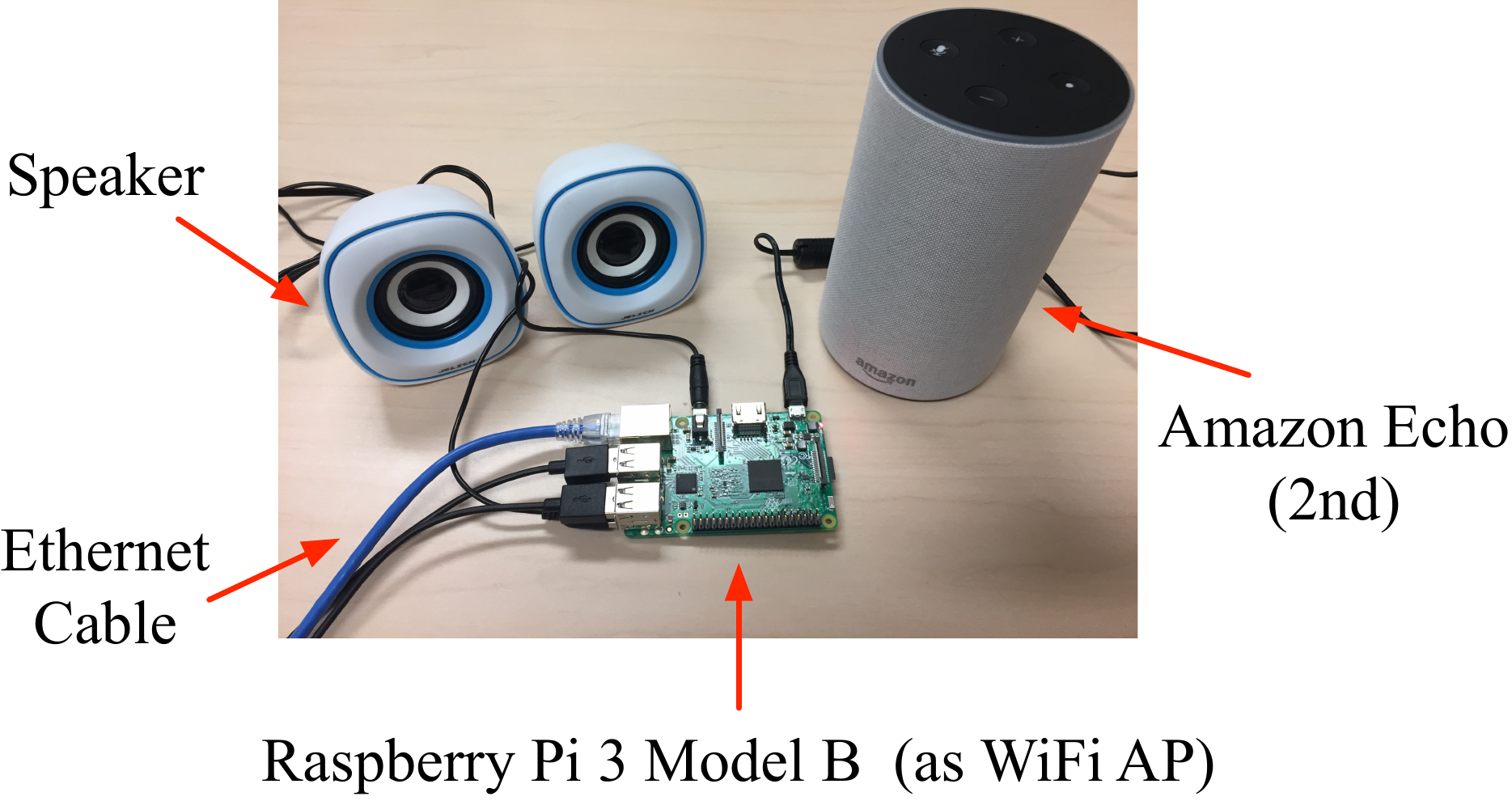}
\vspace{-5pt}
\caption{Our automatic voice traffic collection tool.}
\vspace{-5pt}
\label{fig:crawler}
\end{figure}

To automatically collect traffic traces for the study of voice command fingerprinting, we designed a \textit{voice command traffic collection tool} as shown in Fig.~\ref{fig:crawler}. This tool consists of three main components: a Raspberry Pi, a regular speaker and a smart speaker. The Raspberry Pi is leveraged as a compromised WiFi Access Point, which can capture and store traffic traces. It is connected to the Internet through an Ethernet cable. A smart speaker is connected to the Raspberry Pi through WiFi. The regular speaker is connected to the Raspberry Pi via an audio cable. An Amazon Echo (2nd generation) is used as an example in Fig.~\ref{fig:crawler}. The system is generic, and also works for other smart speakers, such as Google Home. 

We prepared a list of commands in text and utilized text-to-speech APIs to generate audio files of voice commands. Specifically, we leveraged Google Cloud text-to-speech API 
and Amazon Polly  
to generate multiple audio files for each command. 
We utilized multiple different automated voices (in US English), including 3 female voices (Google, Joanna, Salli) and 2 male voices (Joey and Matthew). Google voice is from Google Cloud text-to-speech API and the other four voices are selected from Amazon Polly. For each voice command, our tool generates five audio files, one for each voice. More voices can be supported in this tool if needed. 

We used a Raspberry Pi (Pi 3 Model B) running the Raspian OS. We developed a \texttt{Python} script containing 80 lines of code and run the script on the Raspberry Pi to automatically play audio files one by one. Upon receiving each voice command from the regular speaker, the smart speaker forwards the command to its cloud server and returns a response. A pre-defined time interval is estimated and applied between the start time of two audio files to ensure that the smart speaker can complete each response. \texttt{tcpdump} is executed on the Raspberry Pi to automatically capture traffic traces for each voice command and its response.

\textit{To the best of our knowledge, this is the first automatic tool that can be utilized for encrypted traffic analysis on smart speakers.} Based on our tests, this tool is capable of automatically collecting approximately 3,000 traffic traces per day without any human interaction. We leveraged this tool and collected two large datasets. 
Details of these two datasets are presented in Sec.~\ref{sec:dataset}.

\subsection{Data Format}
A raw traffic trace captured by \texttt{tcpdump} is first converted to a sequence of tuples as described in Eq.~\ref{equation:genetric}. 
Since this format cannot be directly passed to neural networks, we further transformed the data into two different formats. 

\textbf{Binary Format.} Given a traffic trace $T_{C} = \langle (b_{1}, s_{1}, t_{1}), ..., (b_{m}$, $s_{m}$, $t_{m}) \rangle$, its binary format is $T_{C} = (b_{1}, ..., b_{m})$,
which keeps only the direction of each packet.   

\textbf{Numeric Format.} Given a traffic trace $T_{C} = \langle (b_{1}, s_{1}, t_{1}), ..., (b_{m}$, $s_{m}$, $t_{m}) \rangle$, its numeric format is $T_{C} = (b_{1}\times s_{1}, ..., b_{m}\times s_{m})$, 
which keeps direction and packet size. 

For instance, given a traffic trace $T_C = \langle (1, 20, 0.5)$, $(1, 50, 2.1)$, $(-1, 250, 5.3)$, $(1, 100, 6.7)\rangle$, 
its binary format is $(1, 1, -1, 1)$ and its numeric format is $(20, 50$, $-250$, $100)$. 

It is worth mentioning that previous studies in deep-learning-based website fingerprinting often examine binary format only, as Tor networks send fixed-size packets (i.e., \textit{cells}) \cite{SIJW18}. In our study, we observed that the packet size of voice traffic on smart speakers varies, so we examined results in both formats. 
Data in the numeric format is further normalized using MinMaxScaler (with \texttt{scikit-learn} library) before 
being used for neural networks, as the inputs for deep learning should be within the range of $[-1, 1]$. 

Since neural networks require the same input length for different classes, different traffic traces are adjusted to an identical vector size. If the original length is smaller than the uniform vector size, we padded the input with 0s; if it is greater, we trimmed the input by dropping data after the uniform vector size. This uniform vector size is one of the hyperparameters we tuned in our experiments. Padding traces to an identical size is commonly used in other attacks, such as website fingerprinting \cite{SIJW18, RPJGJ18}, if neural networks are used as classifiers.  

\subsection{Neural Networks}
We implement proof-of-concept attacks using three neural networks, including  {Convolutional Neural Networks}, {Long Short-Term Memory} and {Stacked Autoencoder}, respectively. For each type of neural networks, we exploited several different structures, and reported the structure that achieved the highest accuracy. 

\textbf{Convolutional Neural Network (CNN).}
CNN has been widely used in various classification problems. 
According to studies in other research areas, especially image classification \cite{LBH15}, the accuracy of CNN often outperforms other neural networks. It is likely that CNN would produce higher accuracy than others in voice command fingerprinting. 

\textbf{The structure of our CNN.} Our CNN (described in Fig.~\ref{fig:cnn} in Appendix) consists of 11 layers, including 1 input layer, 4 convolutional layers, 5 pooling layers and 1 output layer. 

\textbf{Long Short-Term Memory (LSTM).}
Long Short-Term Memory, is an advanced version of a Recursive Neural Network. It can mitigate the vanishing gradient and exploding gradient problem in (vanilla) RNNs. LSTM performs better over \textit{time-series data}. Encrypted traffic, in essence, is time-series data, which implies LSTM could outperform others in voice command fingerprinting. 

\textbf{The structure of our LSTM.} Our LSTM (illustrated in Fig.~\ref{fig:lstm} in Appendix) consists of 1 input layer, 5 LSTM layers, and 1 output layer. 
Each LSTM layer consists of multiple LSTM units. 

\textbf{Stacked AutoEncoder (SAE).}  
SAE includes an encoder, a code and a decoder. 
SAE first compresses data to a smaller number of dimensions and then produces the output by decoding the compressed data. 
SAE can efficiently extract features from a great number of dimensions and increase accuracy in classification. A traffic trace includes hundreds of packets, where the side-channel information of one packet is a dimension. Leveraging SAE in voice command fingerprinting could attain better attack results. 

\textbf{The structure of our SAE.}  
Our SAE (described in Fig.~\ref{fig:sae} in Appendix) consists of 9 layers, including 4 layers for encoder, 1 layer for code and 4 layers for decoder. Once it is trained, the encoder and the code are extracted and one dense layer is attached to the end in order to perform classification.

\begin{figure}[ht]
\centering
\includegraphics[width=6cm]{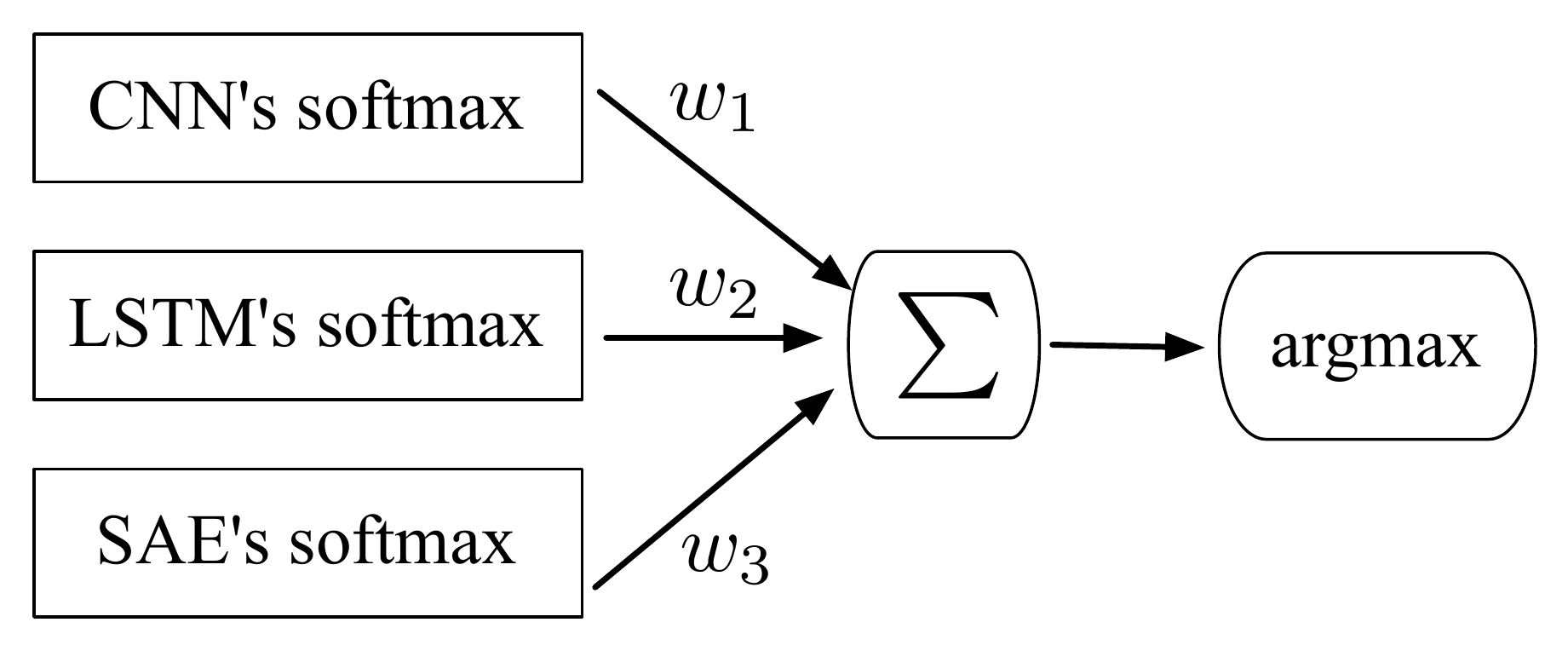}
\vspace{-5pt}
\caption{The main idea of our ensemble learning.}
\vspace{-10pt}
\label{fig:ensemble}
\end{figure}

\textbf{Ensemble Learning.} 
In addition to the three neural networks, we also harness \textit{ensemble learning} \cite{OM99}. 
Ensemble learning combines the predictions of multiple networks, which is known to result in better performance than any single network. The step for ensemble learning in our attack is shown in Fig.~\ref{fig:ensemble}. Ensemble learning takes the output of the output layer (i.e., \texttt{softmax} function) of each single network, calculates a summation and obtains a prediction with an \texttt{argmax} function. 
We assign weights $w_{1}$, $w_{2}$ and $w_{3}$ for the three neural networks. If the weight for each single network is the same, then it is called \textit{average ensemble}; otherwise, it is referred to as \textit{weighted ensemble} \cite{OM99}.

\section{Encrypted Voice Traffic Datasets}
\label{sec:dataset}

\textbf{Overview.} We collected two datasets with the tool described in Sec.~\ref{sec:attack}. We refer to the two datasets as \textit{Amazon Echo Dataset} and \textit{Google Home Dataset}. 
We ran our data collection tool in a 200 square-foot room on campus with a reasonable level of background noise and human activities. Those include a person working on a laptop/PC and typing next to the tool, opening and closing the door occasionally as regular office hours, and students passing the door and the windows of the room from the hallway. The two datasets were collected from March 2019 to August 2019.   

\textbf{Amazon Echo Dataset.} For the closed-world setting, we collected 150,000 encrypted traffic traces on Amazon Echo (2nd generation). This dataset includes 100 voice commands and 1,500 traffic traces per command. This list of commands is referred to as \textbf{\textit{the monitored list}} by following the literature in website fingerprinting \cite{SIJW18, RPJGJ18}. For the open-world setting, we chose another 100 voice commands and collected 200 traffic traces for each class. This list of commands is referred to as \textbf{\textit{the unmonitored list}}. We leveraged 5 different voices (Google, Joanna, Joey, Matt, and Salli) for each command in both the closed-world and the open-world setting. Each voice is associated with 20\% of traffic traces. This effort lasted approximately 8 weeks in total to complete all the traffic traces in the closed-world setting and approximately 3 weeks to finish the traffic traces in the open-world setting of this dataset. The size of this data is 26.05 GBs (closed-world) and 4.07 GBs (open-world). 

The voice commands in our study were selected based on Amazon Echo weekly emails. 
These emails include popular voice commands that Amazon Echo users often ask. We selected voice commands based on Amazon Echo weekly emails from December 2018 to March 2019. 
A full list of those commands can be found at \cite{DeepVCFingerprinting}. 

\textbf{Google Home Dataset.} We collected another dataset on Google Home. For this dataset, we only chose a list of 100 voice commands for the study of the closed-world setting. As 21 voice commands in the monitored list of Amazon Echo dataset did not work on Google Home, we chose another 21 new commands and added them to the monitored list of this dataset. Since the wake word is different, we regenerated the audio files with the same voices. For each command, we still collected 1,500 traffic traces with 20\% traces per voice. This effort lasted 9 weeks and resulted in a dataset of size 33.57 GBs. 

\textbf{Removing Invalid Traffic Traces.}
We removed \textit{invalid} traffic traces due to unexpected errors of a smart speaker. For example, an invalid traffic trace could happen when a voice command was correctly played but there was no response from the server. Removing invalid traffic traces is also common in the data collection of other encrypted traffic analysis, such as website fingerprinting \cite{SMRW19}.  

Only a very small number of traces are invalid and removed. For the Amazon Echo dataset, it has 148,770 valid traffic traces (99.18\%) for the closed-world setting. The minimal number of traces belonging to one class is 1,340, and the maximum number is 1,500. There are 19,953 valid traffic traces (99.77\%) for the open-world setting. The Google Home dataset has 149,745 valid traffic traces (99.83\%) for the closed-world setting. 

\textbf{Categories of Voice Commands.} Based on the responses of each command, we grouped commands into three categories, referred to as \textit{single response commands}, \textit{time-sensitive response commands}, and \textit{multiple response commands}. 

A single response command indicates that the response was always (or almost) the same during our data collection. For example, for voice command ``\textit{Where is Mount Rushmore?}" the response from the Amazon server was always the same in our study. 

A time-sensitive response command implies that the response changed overtime. For example, ``\textit{What is the weather today?}" the Amazon server replied a different answer each day. 

A \textit{multiple response} command suggests that we received a number of different responses. However, the content of each response did not change over time. For instance, for the command ``\textit{Tell me a barbecue joke.}" the Amazon server randomly returned one of five possible jokes during our data collection. 

The ratio of each command category of the monitored list in our Amazon Echo dataset is elaborated in Table.~\ref{table:categories}. We further discuss the impact of categories on the attack results in the next section.

\begin{table}[htbp]
\vspace{-5pt}
\small
	\centering  
	\caption{Ratios of Voice Commands in The Three Categories (Amazon Echo Dataset, The Monitored List)}
	\label{table:command:category}
	\begin{tabular}{ccc}  
		\hline
 Single	& Time-Sensitive &  Multiple \\ \hline 
	 $45\%$  & $21\%$ & $34\%$  \\ \hline 
	\end{tabular}
	\vspace{-5pt}
\label{table:categories}
\end{table} 

\begin{figure}[t]
     \centering
  \begin{subfigure}[b]{0.32\textwidth}
    	 \centering
         \includegraphics[width=\textwidth]{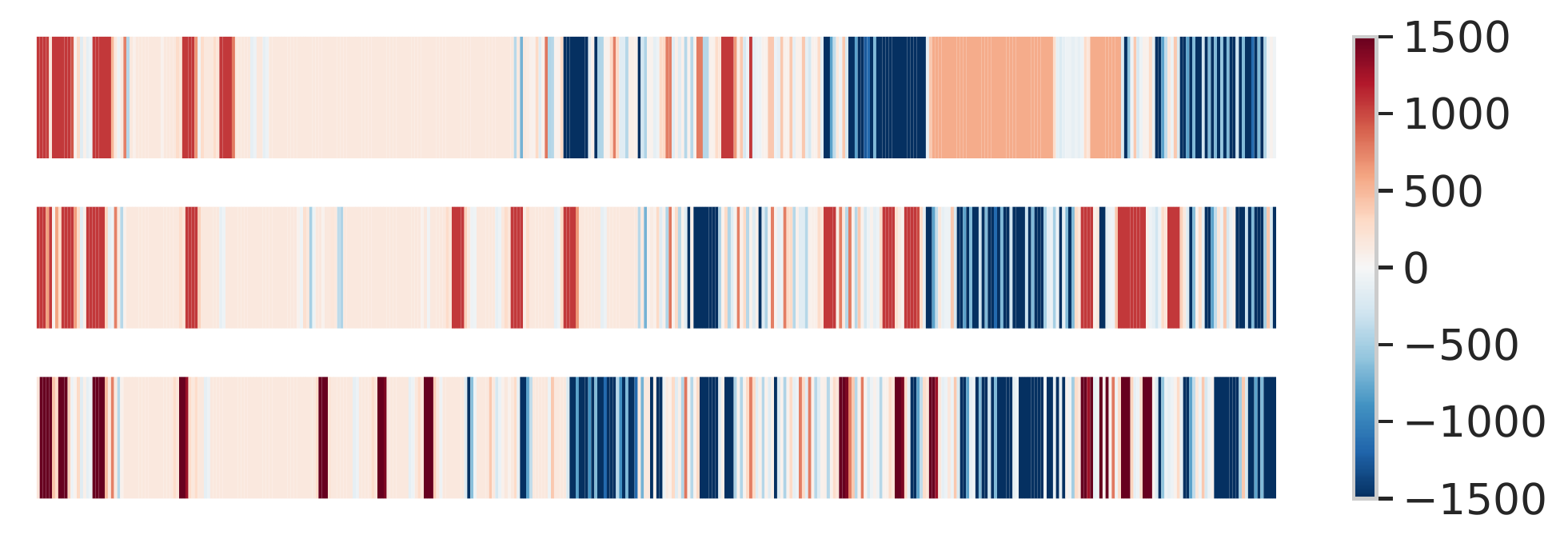}
         \caption{What is my sports update?}
     \end{subfigure}
     \hfill
     \begin{subfigure}[b]{0.32\textwidth}
     	\centering
         \includegraphics[width=\textwidth]{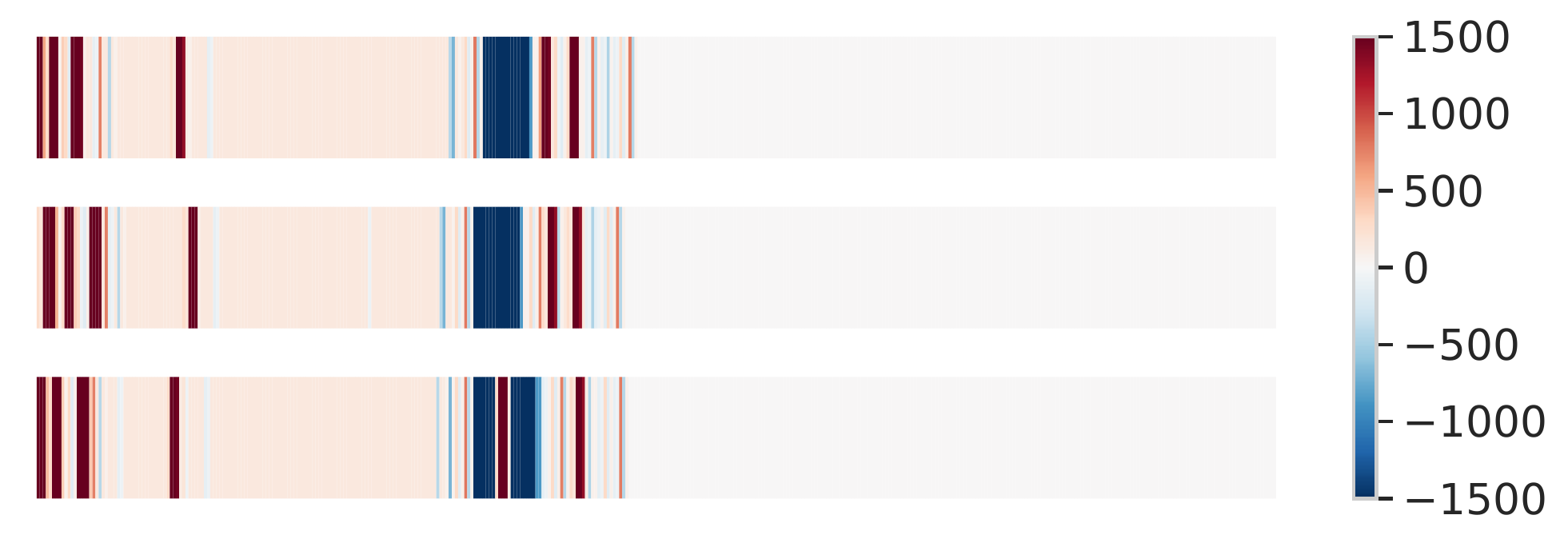}
         \caption{What is the date tomorrow?}
     \end{subfigure}
     \hfill
     \begin{subfigure}[b]{0.32\textwidth}
         \centering
         \includegraphics[width=\textwidth]{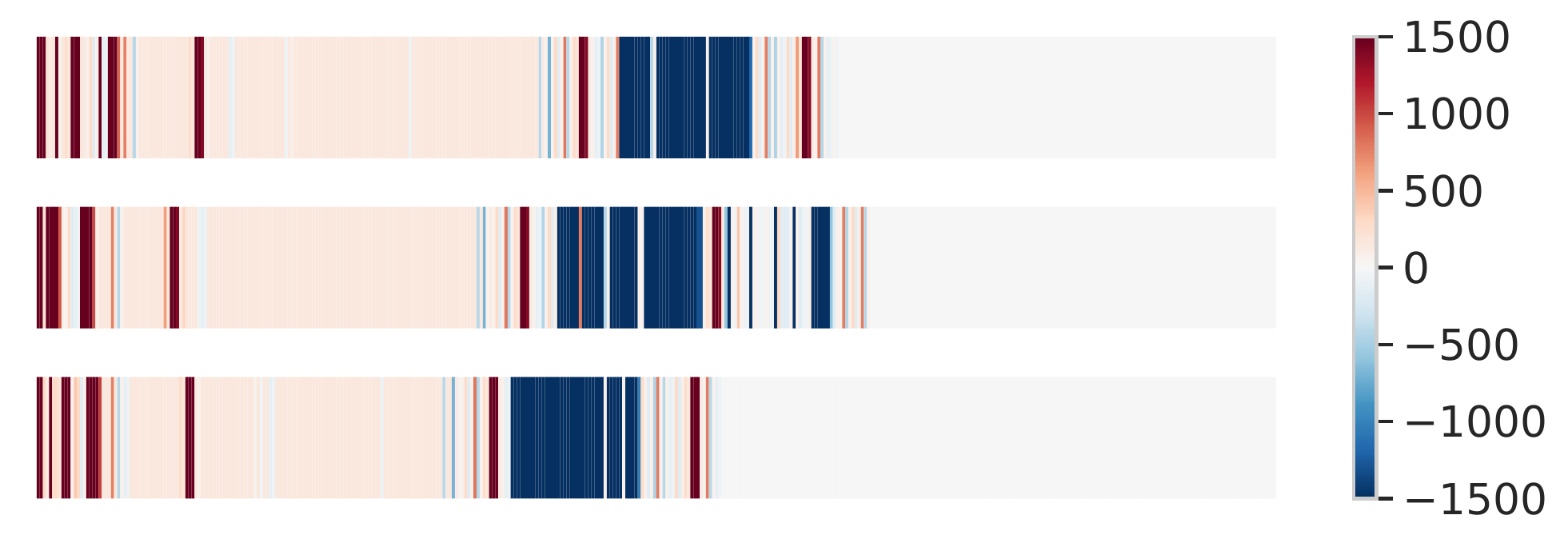}
         \caption{What is my traffic report?}
     \end{subfigure}
     \hfill
     \begin{subfigure}[b]{0.32\textwidth}
         \centering
         \includegraphics[width=\textwidth]{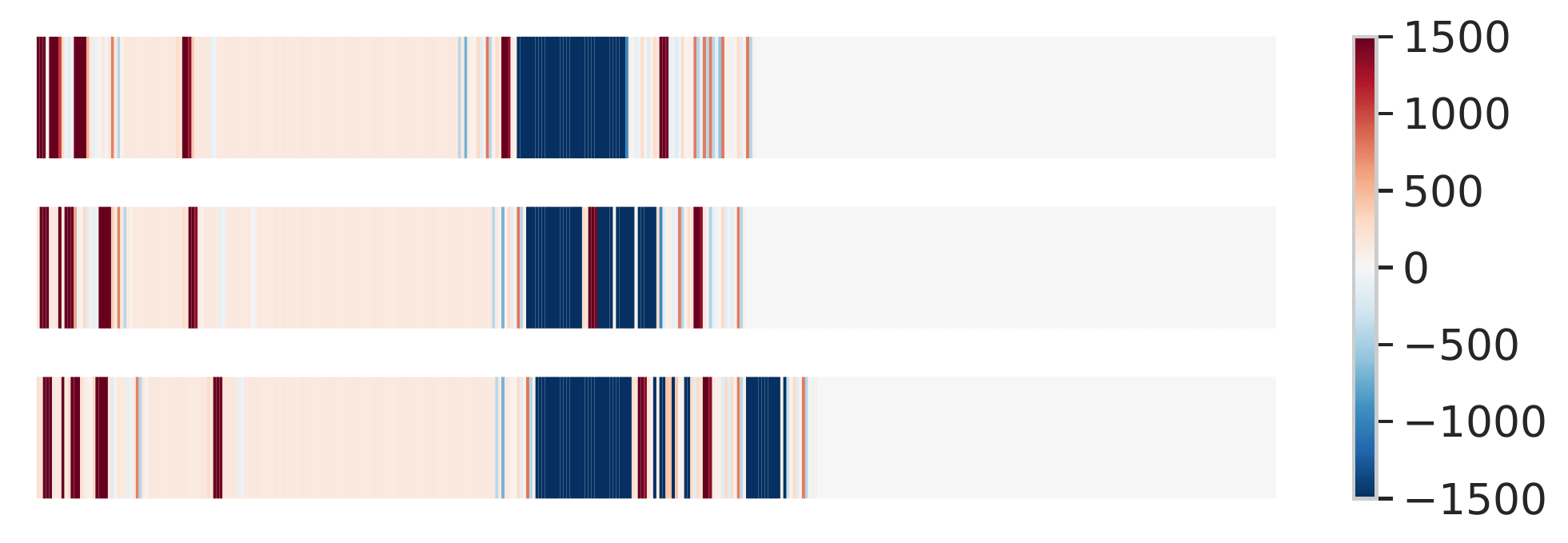}
         \caption{What is the price of Bitcoin?}
     \end{subfigure}
     \vspace{-5pt}
     \caption{Each heat map includes 3 encrypted traffic traces of one voice command. Traffic traces were generated by an Alexa Echo. Outgoing traffic are marked with red color and incoming traffic are marked with blue color. A darker color indicates a packet has a greater packet size.}
     \label{fig:traffic:pattern}
     \vspace{-15pt}
\end{figure}

\textbf{Data Visualization.} Before we evaluated the results of our neural networks, we first visualized some traffic traces from Amazon Echo dataset by generating heat maps of traffic traces from each command. The main purpose of this step is to demonstrate that it is feasible to fingerprint voice commands based on encrypted traffic of smart speakers. Due to space limitation, we only present the heat maps of 4 voice commands in Fig.~\ref{fig:traffic:pattern} and each heat map only contains 10 traffic traces in Amazon Echo dataset.

First, we observe that it is indeed viable to infer voice commands based on encrypted traffic. Specifically, the traffic traces of one command are not exactly the same in each heat map, but are very similar. In addition, the traffic pattern of some commands are completely different. For instance, the traffic traces from the first three heat maps, including heat map (a), (b) and (c), are obviously distinguishable. On the other hand, we also notice that it is not trivial to distinguish all the commands based on heat maps as the traffic pattern of some commands are similar. For example, the difference of traffic traces between (c) and (d) in Fig.~\ref{fig:traffic:pattern} are not obvious. In fact, most of the classes we investigate have similar pattern as heat map (c) and (d). \textit{This creates the need for sophisticated neural networks}

\section{Attack Evaluation}
\label{sec:evaluation:attacks}

\textbf{Experiment Setting.} We implement our proof-of-concept attack in \texttt{Python}. 
We used \texttt{Keras} as the front end and \texttt{Tensorflow} as the back end to implement neural networks. 
We ran our experiments on a Linux machine with Ubuntu 18.04 OS, 2.8 GHz CPU, 16 GB Memory, and a GPU (NIVIDA GeForce GTX 1070). 
We ran 5-fold cross validation. We used 64\% of the data for training, 16\% for validation, and 20\% for testing. 

\textbf{Hyperparameter Tuning.} We searched hyperparameters for neural networks with NNI (Neural Network Intelligence), a Microsoft open-source toolkit. 
We used TPE (Tree-structured Parzen Estimator) as our search/tuning algorithm, which is one of the tuning algorithms provided by NNI.

For each neural network, we ran NNI with 50 iterations at most or stopped the search if it took longer than 200 hours. After we found a set of tuned hyperparameters with the highest accuracy score, we recorded the tuned hyperparameters and trained our networks using these optimal hyperparameters. When we trained each network, we limited the number of training epochs to 500 or stopped the training early if the accuracy did not continue to improve after 10 consecutive epochs. 
The search space and tuned hyperparameters can be found in Appendix.

\begin{table*}[htbp]
\small
	\centering  
	\caption{Attack Results in the Closed-World Setting Based on Both Outgoing and Incoming Traffic (AE: Averaging Ensemble; WE: Weighted Ensemble; ACC: Accuracy; VAR: Variance)}
	\begin{tabular}{cccccccccc}  
		\hline
		
    Dataset 	                  & Format	  & \multicolumn{2}{c}{CNN}	                    & \multicolumn{2}{c}{LSTM}                      &  \multicolumn{2}{c}{SAE}              & AE        & WE    \\  
                                  &           & ACC & VAR                                   & ACC & VAR                                     & ACC & VAR                             & ACC       & ACC \\ \hline
	\multirow{2}{*}{Amazon Echo}  & Binary    & 75.88\%  & $0.56 \times 10^{-5}$            & \textbf{77.51\%} &  $4.68 \times 10^{-5}$     & 66.59\%  &  $0.44 \times 10^{-5}$     & 85.49\%   & \textbf{85.70\%} \\  
	                              & Numeric   & \textbf{89.05\%} & $1.50 \times 10^{-5}$    & 88.65\%  & $0.49 \times 10^{-5}$              & 75.98\% & $0.48 \times 10^{-5}$       & 89.41\%   & \textbf{92.89\%} \\ \hline
    \multirow{2}{*}{Google Home} & Binary     & 95.17\% & $1.62 \times 10^{-5}$             & \textbf{96.90\%} & $2.48 \times 10^{-5}$      & 90.20\% & $0.47 \times 10^{-5}$       & --        & -- \\ 
	                             & Numeric    & \textbf{99.22\%} & $0.06 \times 10^{-5}$    & 98.62\% & $0.13 \times 10^{-5}$               & 92.34\% & $0.28 \times 10^{-5}$       & --        &  --  \\  
\hline
	\end{tabular}
\label{table:attack:format}
\end{table*} 

\begin{table*}[htbp]
\small
	\centering  
	\caption{The Comparison with Previous Methods in the Closed-World Setting Based on Both Outgoing and Incoming Traffic of Amazon Echo Dataset}
	\begin{tabular}{ccccccc}  
		\hline
	Attack Method	            & CNN	            & LSTM          &  SAE      &  CUMUL \cite{PLZHPWE16} & CNS19 \cite{KLWLWS19}   & Random Guess  \\ \hline 
	Accuracy  	                & \textbf{89.05\%}  & 88.65\%       & 75.98\%   & 61.44\%                 & 76.32\%                 & 1\%           \\  
	Training Time (second)      & 5,327             & 23,967        & 1,800     & 6,073                   & 4,421                   & N/A            \\ \hline 
	\end{tabular}
\label{table:attack:comparison}
\end{table*} 

\subsection{Closed-World: Outgoing \& Incoming Traffic}
\label{subsec:test:entire}

We first evaluate attack results in the closed-world setting leveraging both outgoing and incoming traffic. 

\textbf{Which Input Format is More Effective?} 
We first compared the attack results between the binary format and numeric format for each model. As shown in Table~\ref{table:attack:format}, {\textit{the accuracy of the numeric format is much higher than the accuracy of binary format}}. This indicates that data in the numeric format leaves more identifiable fingerprints in the encrypted traffic. 

\textbf{Which Neural Network is More Effective?} We observed that CNN and LSTM resulted in very similar results and were both significantly higher than SAE in the closed-world setting. 
The variance of accuracy in each network is small in 5-fold cross validation, which suggests the neural  networks are stable. 

The results on Google Home dataset indicated similar observations as the ones derived from Amazon Echo dataset. 
For the attacks on Google Home dataset, we used the same hyperparameters for each neural network as the ones in Amazon Echo dataset. This suggests that our neural networks are \textit{transferable} across encrypted traffic from different smart speakers. 

\textbf{Can Ensemble Learning Improve Accuracy?} As presented in Table~\ref{table:attack:format}, our results demonstrate that ensemble learning can improve the attack accuracy nearly 4\% and outperforms each single network in the closed-world setting on Amazon Echo dataset. 

For weighted ensemble, we calculated the normalized weights using the accuracy on the validation data and reported the attack accuracy based on test data. The normalized weights we derived for CNN, LSTM and SAE were 0.35, 0.35, and 0.30 respectively on Amazon Echo dataset. We did not run 5-fold cross validation in ensemble learning as the variance of each single network was very low. For Google Home dataset, as CNN already achieved extremely high accuracy with over 99\%, we did not use ensemble learning.   

\textbf{Comparison with Previous Studies.}
We compared our results with previous studies. Particularly, we compared our neural networks running the numeric format with two conventional machine learning attack methods, CUMUL \cite{PLZHPWE16} and CNS19 \cite{KLWLWS19}, on Amazon Echo dataset in the closed-world setting. 
We chose CUMUL as it is one of the most effective attack methods, and its accuracy is comparable with deep-learning-based methods in website fingerprinting \cite{RPJGJ18}. 
CNS19 \cite{KLWLWS19} manually selected a feature set and implemented the classifier with AdaBoost. 
We implemented both CUMUL and CNS19 with \texttt{Python} in our comparison. 

As shown in Table \ref{table:attack:comparison}, CNS19 achieved 76.32\% accuracy on our Amazon Echo dataset in the closed-world setting, which is significantly higher than the accuracy of 33.8\% reported in \cite{KLWLWS19}. Since we applied the same feature set and same classifier as CNS19, this accuracy increase is likely because the size of our dataset is significantly greater than the dataset utilized in CNS19. Specifically, our dataset has 1,500 traces per class while the dataset in CNS19 only has 10 traces per class. 
CNS19 outperformed CUMUL and SAE but was outperformed by our CNN and LSTM in the comparison. 

\begin{table*}[htbp]
\small
	\centering  
	\caption{Attack Results in the Closed-World Setting Based on Incoming Traffic Only (AE: Averaging Ensemble; WE: Weighted Ensemble; ACC: Accuracy; VAR: Variance)}
	\begin{tabular}{cccccccccc}  
		\hline
		
                Dataset          & Format	 & \multicolumn{2}{c}{CNN}	            & \multicolumn{2}{c}{LSTM}                  &  \multicolumn{2}{c}{SAE}                  & AE                & WE    \\  
                                 &           & ACC & VAR                            & ACC & VAR                                 & ACC & VAR                                 & ACC               & ACC \\ \hline
	\multirow{2}{*}{Amazon Echo} & Binary    & 24.40\% & $0.18 \times 10^{-5}$      & 24.38\%  & $0.28 \times 10^{-5}$          & 24.65\% & $1.72 \times 10^{-5}$           & 24.44\%           & 24.16\% \\  
	                             & Numeric   & 81.69\%  & $1.70 \times 10^{-5}$     & \textbf{85.09\%} &  $1.93 \times 10^{-5}$          & 73.77\%  &  $2.37 \times 10^{-5}$         & 84.41\%           & \textbf{86.09\%} \\  \hline
    \multirow{2}{*}{Google Home} & Binary    & 8.90\%  & $0.88 \times 10^{-5}$      & 9.25\% & $0.21 \times 10^{-5}$            &8.92\% & $0.35 \times 10^{-5}$             & 9.26\%            & \textbf{9.35\%}   \\  
	                             & Numeric   & 88.50\% & $6.97 \times 10^{-5}$      & \textbf{92.24\%} & $7.84 \times 10^{-5}$           & 81.57\%  &  $3.01 \times 10^{-5}$         & 91.66\%           & \textbf{92.48\%}  \\  
\hline
	\end{tabular}
\label{table:attack:onlyIncoming}
\end{table*}

\textbf{The Impact of The Number of Traces.}
Next, we evaluated attack accuracy with different sizes of data. Specifically, we kept the same 100 commands in the monitored list of Amazon Echo dataset, but we randomly selected a subset of traffic traces from each command based on a given number of traces per class. We increased the number of traces per class from 100 to 1,300 with an interval of 100. We tested attack accuracy of five methods, including CNN, LSTM, SAE, CNS19, and CUMUL for each different size. For different sizes, we used the same hyperparameters, retrained the neural networks each time based on the corresponding data. 

\begin{figure}[t]
\centering
\includegraphics[width=6cm]{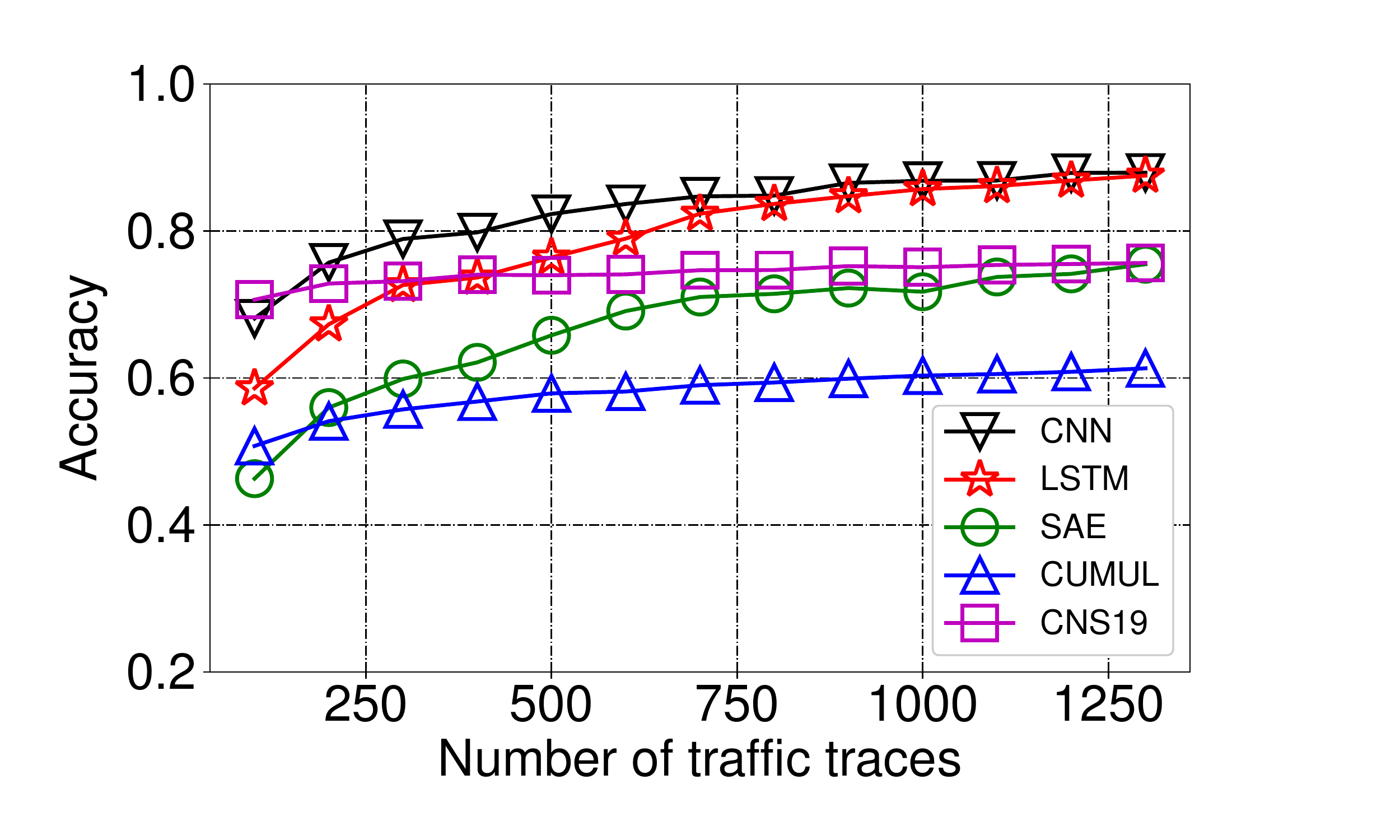}
\vspace{-10pt}
\caption{The impact of data size on attack accuracy.}
\vspace{-15pt}
\label{fig:test:num_test}
\end{figure}

As shown in Fig.~\ref{fig:test:num_test}, for neural networks, we observed significant improvements by utilizing a greater number of traffic traces. 
CNN performed the best among all the three deep learning models with every size we tested. 
CNS19 outperformed others only when the number of traces for each class was 100. 
The accuracy of CNS19 and CUMUL increased slowly and did not gain much improvement when the number of traces per class approached 1,300. 

We could obtained more fine-grained results in Fig.~\ref{fig:test:num_test} if we increase the data size with a smaller interval (e.g., 10). In that case, we would need to re-train hundreds of neural networks, which is time-consuming even with a GPU. Using the interval of 100 was sufficient for us to observe the impact of data size on attack accuracy.

\subsection{Closed-World: Incoming Traffic Only}
\label{subsec:test:onlyIncoming}

\textbf{Will Our Attacks Be Effective on Human Voices?} Our attack results using both outgoing and incoming traffic are promising. However, a key question we have not investigated is whether our attacks will be effective on human voices in the real world. 

The encrypted traffic traces in our two datasets were triggered with five different automated voices from public Text-to-Speech APIs. Although these automated voices render a certain degree of variance in voices, human voices vary significantly due to multiple factors, including gender, age, and accent. These factors could lead to high variances in the voice data of the same voice command, which may change the pattern of the traffic to the server and affect the accuracy of our attacks.

In addition, humans can ask the same intent with variations in text. For example, a human can ask \textit{"Will it rain tomorrow?"} or \textit{"Is it going to rain tomorrow?"}, where both commands have the same semantic intent and will receive the same response. These variations in text could also affect in the pattern of outgoing traffic.

While investigating all the variances in human voices and texts is obviously challenging (and nearly impossible), \textit{\textbf{one of our key observations is that all these variances do not affect the incoming traffic from the server.}} Specifically, humans can ask the same commands with different voices and texts, but as long as the AI-based voice services on the server side understand correctly, the responses (as well as the incoming traffic) are not affected by these variances in voices and texts. Thus, we further conducted experiments using incoming traffic only to prove our attacks would be effective in the real world.

Our results in Table~\ref{table:attack:onlyIncoming} shown that our neural networks are effective even considering incoming traffic only. For instance, LSTM still achieved 85.09\% accuracy with the numeric format 
on Amazon Echo dataset. Between the numeric format and binary format, we observed that the numeric format were more effective.   

For each neural network, we re-tuned hyperparameters based on incoming traffic only. The tuned hyperparameters can be found in Appendix. Our results based on incoming traffic only also indicated that the primary reason causing identifiable voice commands over encrypted traffic on smart speakers is likely because their AI-based voice services response in a deterministic or predictable manner, which leave distinguishable fingerprints in encrypted traffic. 

\textbf{The Impact of Different Voice Command Categories.} 
We also evaluated the impact of the categories on attack accuracy using incoming traffic only. We separated Amazon Echo dataset into three subdatasets based on the categories of each command. We evaluated attack accuracy based on each subdataset. 

According to the results, single response commands and time-sensitive response commands were easier to infer. For instance, CNN achieved 87.57\% accuracy for single response commands and 88.94\% accuracy for time-sensitive response commands. For commands with multiple responses, CNN still revealed significant private information with over 75\% accuracy. 

\begin{table}[tbp]
\small
	\centering  
	\caption{Attack Results for Different Categories in the Closed-World Setting (Incoming Traffic, Numeric Format)}
	\begin{tabular}{lccc}  
		\hline
		                & CNN	    & LSTM      & SAE           \\ \hline 
	Single  	        & 87.57\%   & 81.50\%   & 80.92\%       \\ 
	Time-Sensitive      & 88.94\%   & 86.67\%   & 83.95\%       \\ 
	Multiple            & 75.92\%   & 74.46\%   & 68.41\%       \\ \hline
	\end{tabular}
	\vspace{-5pt}
\label{table:attack:type}
\end{table}

\subsection{Open-World: Incoming Traffic Only} 
We evaluated the open-world setting with our Amazon Echo dataset. To keep the data balanced, we used 200 valid traces per class in the monitored list and we used all the valid traces of each class in the unmonitored list. We retrained each neural network under the assumptions of the open-world scenario, which is a binary classification with the aim to decide whether or not a traffic trace is associated with the monitored list. We only reported the results in the open-world setting with incoming traffic only. 

\begin{table}[htbp]
\small
\vspace{-5pt}
	\centering  
	\caption{Attack Results in The Open-World Setting on Amazon Echo (Incoming Traffic Only)}
	\begin{tabular}{llcccc}  
	\hline
	Format                      & Metric    & CNN       & LSTM       & SAE              & AE                    \\ \hline 
	\multirow{3}{*}{Numeric}    & ACC       & 99.94\%   & 100\%      & 99.92\%          & \textbf{100\%}      \\ 
	                            & TPR  	    & 100\%     & 100\%      & 99.93\%          & \textbf{100\%}        \\ 
	                            & FPR       & 0.12\%    & 0.00\%     & 0.08\%           & \textbf{0.00\%}       \\ 
	\multirow{3}{*}{Binary}     & ACC       & 57.09\%   & 57.56\%    & 50.54\%          & 56.33\%       \\ 
	                            & TPR  	    & 66.04\%   & 56.46\%    & 47.41\%          & 57.31\%       \\  
	                            & FPR       & 51.98\%   & 41.32\%    & 46.28\%          & 44.67\%                   \\  \hline 
	\end{tabular}
	\vspace{-5pt}
\label{table:attack:open:world:all}
\end{table}

Our results in Table~\ref{table:attack:open:world:all} show that, with data in numeric format, an attacker can decide whether a traffic trace is associated with the monitored list with an extremely high true positive rate and a very low false positive rate.

\section{A Defense against Fingerprinting} 
\label{sec:defense}

We present a proof-of-concept defense to mitigate the privacy leakage against voice command fingerprinting. It integrates two existing primitives, including adaptive padding \cite{SW06} and differential privacy \cite{XRZ15}, to obfuscate traffic pattern. 

\textbf{Defense Details.} 
To minimize latency, 
we first deploy adaptive padding in our defense. Adaptive padding, which was proposed in \cite{SW06}, adds dummy packets and introduces no latency. Dummy packets are inserted based on the distribution of interarrival time and each real packet is still sent at the original timestamp. As a result, it hides traffic bursts and traffic gaps. Details of adaptive padding can be found in \cite{SW06}. This primitive has been used in WTF-PAD \cite{JIPDW16} as a defense in website fingerprinting. However, adaptive padding does not hide other traffic fingerprints, such as traffic length or packet size. Recent studies \cite{SIJW18} have shown that \textit{leveraging adaptive padding alone} is not effective against deep-learning-based attacks. 

To maintain efficacy, we further obfuscate fingerprints that are not well protected by adaptive padding. First, we randomly determine the size of dummy packets based on the distribution of real packet size. Second, we extend the length of different traffic traces to identical obfuscated traffic length. Specifically, after sending the last real packet in each trace, our defense will keep producing dummy packets with adaptive padding until it reaches an obfuscated traffic length. Instead of padding all traffic traces to the same length, which is less efficient, our defense extends a trace with the traffic length of $m$ (i.e., the total number of packets) to an obfuscated traffic length of $m'$, where $2^{a-1}<m\leq m'= 2^{a}$ and $a$ is an integer. 

\begin{table*}[htbp]
\small
	\centering  
	\caption{Defense Results in The Closed-World Setting among Different Methods with $\epsilon$={0.005}}
	\begin{tabular}{lcccccc}  
		\hline
		                                & {CNN}	    & {LSTM}      & SAE       & AE                &  CUMUL    & CNS19     \\ \hline 
	No Defense  	                    & 89.05\%   & 88.65\%   & 75.98\%   & \textbf{89.41\%}  & 61.44\%   & 76.32\%   \\ 
	Training with original traffic      & \textbf{1.23\%}    & 1.05\%    & 1.12\%    & 1.07\%            & 1.97\%    & 1.77\% \\  
	Training with obfuscated traffic    & 26.81\%   & 19.48\%   & 15.69\%   & \textbf{28.42\%}  & 17.14\%   & 14.59\%   \\ \hline 
	\end{tabular}
	\vspace{-5pt}
\label{table:defense}
\end{table*} 

\begin{figure}[ht]
\centering
\includegraphics[width=7cm]{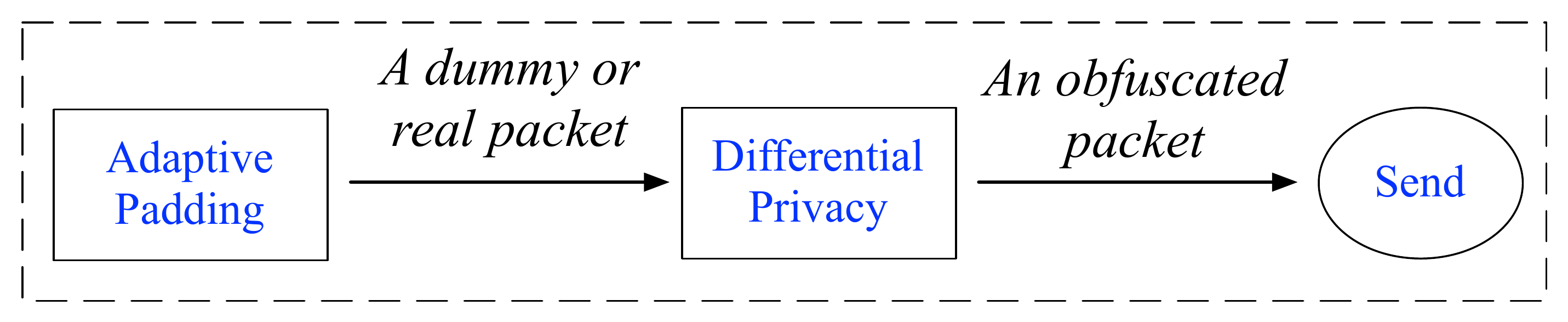}
\caption{Our defense obfuscates each packet on the fly.}
\vspace{-10pt}
\label{fig:defense}
\end{figure}

Third, our defense applies differential privacy to obfuscate packet size on the fly. Specifically, we leverage $d^{*}$-privacy \cite{XRZ15} to add noise to modify packet size, where $d^{*}$-privacy is a variation of differential privacy on time-series data. With $d^{*}$-privacy, the obfuscated outputs of two identical length sequences with a distance of $d$ are indistinguishable. Due to space limitation, details of $d^{*}$-privacy can be found in \cite{XRZ15}. A recent study \cite{ZHRZ19} has demonstrated that $d^{*}$-privacy is effective in obfuscating one-way encrypted traffic in video streams. Building upon \cite{ZHRZ19}, we utilize this primitive to obfuscate bi-directional traffic in our study. A high-level description of our defense of obfuscating each packet size is described in Fig.~\ref{fig:defense}. 

Given a (real or dummy) packet, if noise produced by $d^{*}$-privacy is positive (e.g., $\sigma$), then additional dummy data is inserted to increase the packet size (e.g., $l' = l + \sigma$); if noise is negative (e.g., $-\sigma$), then the packet size will be reduced (e.g., $l' = l - \sigma$) and a corresponding portion of a packet (e.g., $\sigma$) will be buffered until a subsequent real or dummy packet is available in a traffic trace. We implement the buffer as a queue, which sends buffered data before sending new data. 


Unlike \cite{ZHRZ19}, which operates $d^{*}$-privacy (or differential privacy in general) over \textit{bins}, our defense adds the noise on packets. A bin consists of packets within a fixed-size interval. It serves better for one-direction traffic in \cite{ZHRZ19}. For bi-directional traffic, one time interval may have traffic on both directions, which makes it hard to aggregate as one bin and apply noise to traffic. Adding noise directly on packets is more suitable for the bi-directional traffic in our problem. Besides, this minimizes latency, as our defense does not need to buffer the packets in each bin before inserting noise. 

\textbf{Discussions.} Adaptive padding complements differential privacy in two aspects. First, it hides traffic bursts and traffic gaps, which differential privacy alone does not. Second, for buffered data caused by negative noise, the dummy packets produced by adaptive padding can send buffered data sooner, which minimizes latency. 

\textbf{Assumptions on Defense.} We apply our defense on both incoming and outgoing traffic to minimize the privacy leakage from the encrypted traffic. While our results in the previous section showed that incoming traffic plays  a dominating role in the attacks, it is still necessary to obfuscate outgoing traffic to preserve privacy against attacks based on traditional machine learning algorithms, such as CUMUL and CNS19. 

We assume the server will obfuscate incoming traffic, and the smart speaker (or a proxy) will obfuscate outgoing traffic. In practice, the server can calculate the distribution information of interarrival time and packet size, and other information that are required to perform the defense, and forward these information to a smart speaker. 
As we do not have the capability to change the current network protocol, we run simulations of our defense to generate obfuscated traffic from real traffic and demonstrate its efficacy. 


\section{Defense Evaluation}
\label{sec:evaluation:defense}

We implemented our defense in \texttt{Python}. We produced obfuscated traffic traces based on Amazon Echo dataset. The distribution of interarrival time and packet size we used in adaptive padding are generated based on Amazon Echo dataset. For differential privacy, we generated multiple versions of obfuscated datasets based on different values of privacy parameter $\epsilon$. Privacy parameter $\epsilon$ decides the privacy protection (i.e., the noise level) rendered by differential privacy. A smaller value of $\epsilon$ generates higher noise produced and offers stronger privacy protection. 

We assessed the performance of the defense in two cases: (1) \textbf{\textit{Training with original traffic}}. In this case, neural networks are trained based on original traffic traces, but test data are obfuscated; (2) \textbf{\textit{Training with obfuscated traffic.}} In this scenario, we assume an attacker adapts to the defense, where it trains neural networks with obfuscated traffic traces and tests with obfuscated traffic traces. 

\begin{figure}[t]
\centering
\includegraphics[width=7cm]{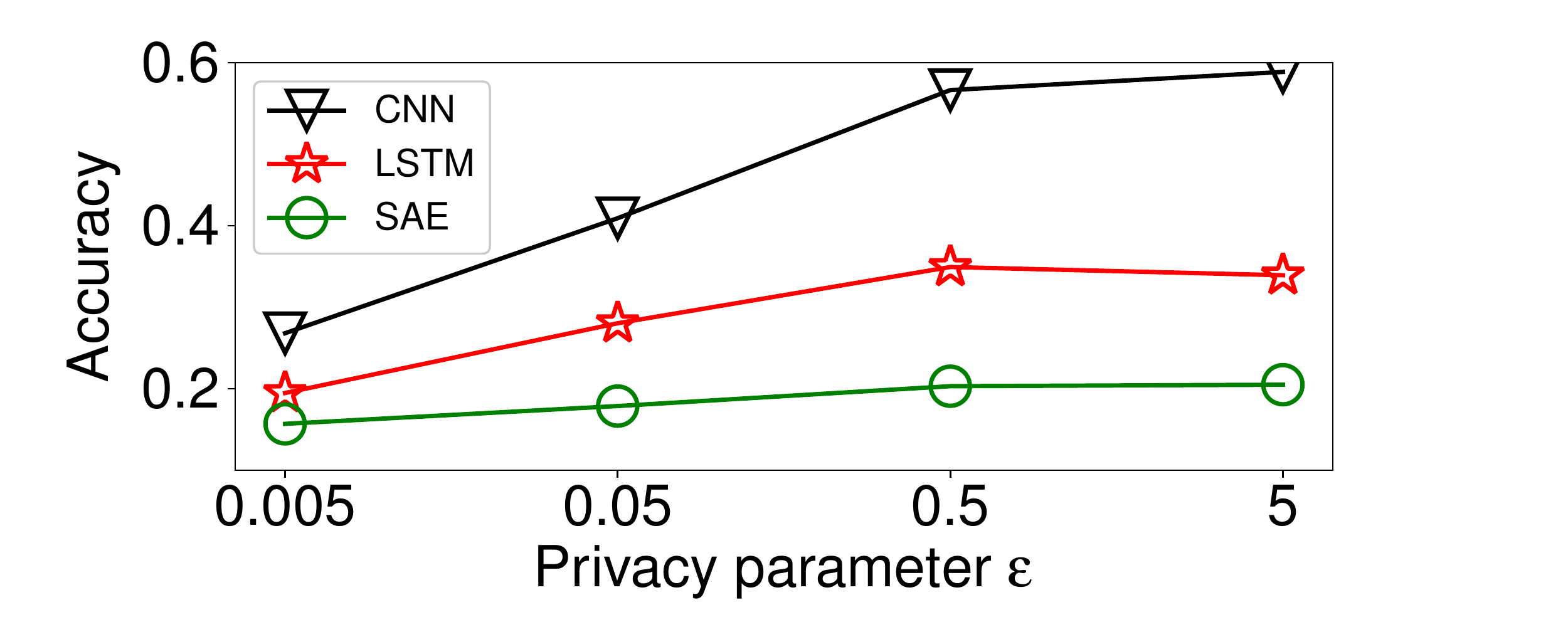}
\vspace{-10pt}
\caption{The impact of privacy parameter on defense with inputs in the numeric format.}
\label{fig:defense:accuracy:numeric}
\centering
\includegraphics[width=7cm]{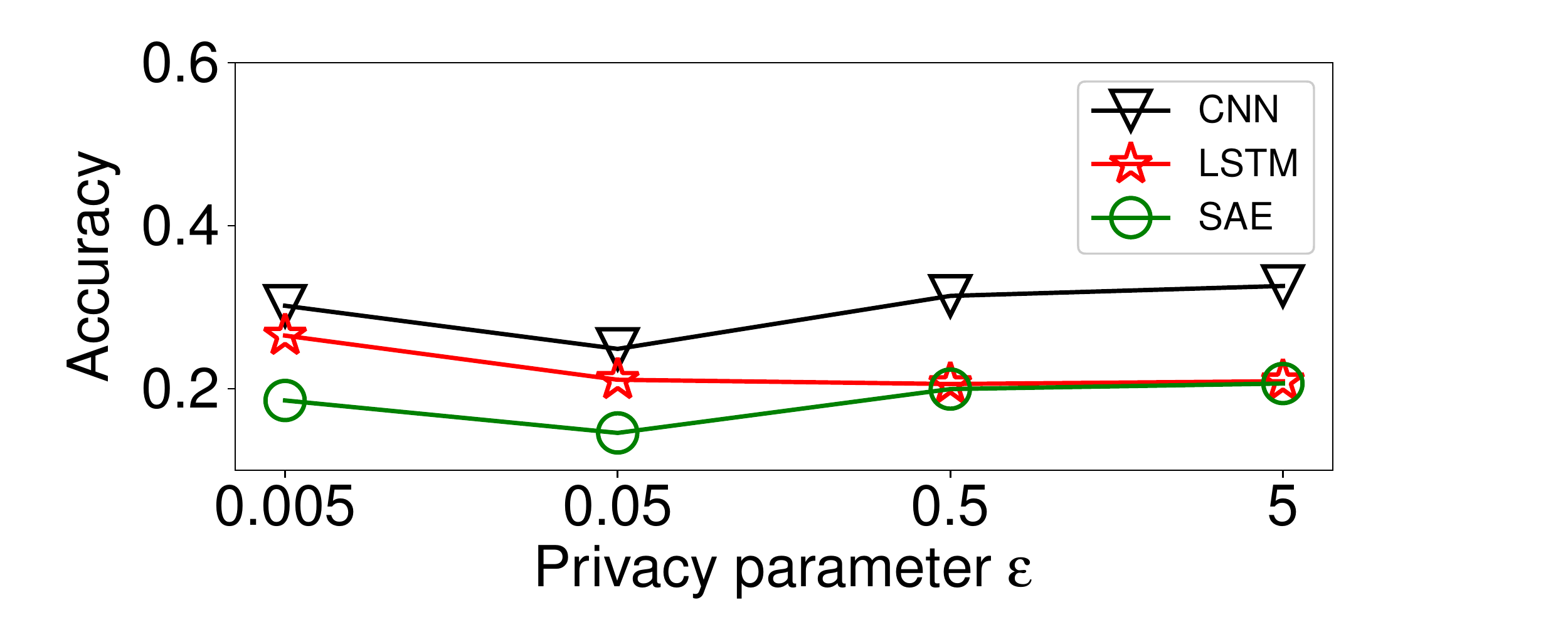}
\vspace{-10pt}
\caption{The impact of privacy parameter on defense with inputs in the binary format.}
\vspace{-10pt}
\label{fig:defense:accuracy:binary}
\end{figure}

As shown in Table \ref{table:defense}, when training with original traffic, our defense can suppress attack accuracy of CNN to 1.23\% in the closed-world setting, which is nearly the same as random guess. If an attacker trains a CNN with obfuscated traffic traces, it can improve its accuracy back to 26.81\%, which is still significantly lower compared to the accuracy without defense. With average ensemble, an attacker can attain 28.41\% accuracy (v.s. 89.41\% with no defense). Our defense is also effective against CUML and CNS19. 

\textbf{The Impact of Privacy Parameter.} In Fig.~\ref{fig:defense:accuracy:numeric}, we show the impact of $\epsilon$ on attack accuracy where the inputs are in the numeric format. 
When $\epsilon$ decreases, the noise level generated by differential privacy increases, which can reduce attack accuracy.  

We also studied the impact of privacy parameter $\epsilon$ on attack accuracy with inputs in the binary format. As illustrated in Fig~\ref{fig:defense:accuracy:binary}, we found that attack accuracy remained relatively stable when we changed privacy parameter. The reason is that changing privacy parameter $\epsilon$ does not have effect on the inputs if they are in the binary format. There are still some minor changes in accuracy as the privacy parameter changes in Fig.~\ref{fig:defense:accuracy:binary}. It is because the obfuscated datasets in binary format are not exactly the same across different values of $\epsilon$ as adaptive padding is a probabilistic algorithm. 

It is worth mentioning that when $\epsilon=0.005$, \textit{accuracy with inputs in the binary format is slightly higher than the one in the numeric format}. For instance, CNN achieved 30.18\% accuracy compared to 26.81\% in the numeric format. Average ensemble attained \textbf{32.18\%} accuracy compared to 28.42\% in the numeric format. 

\textbf{Latency.} 
We examined the latency of our defense by evaluating \textit{latency per packet} and also \textit{latency per traffic trace}. Latency per packet indicates how many milliseconds it takes to clear the (potential) buffered data for each real packet. Latency per traffic trace suggests how many extra milliseconds it takes to complete sending all the real packets of a traffic trace. 
Our results in Table~\ref{table:defense:overhead} show that the defense introduced minimal latency. 
The latency per packet hardly aggregated over packets as the buffered data is cleared rather soon either by dummy packets or the next real packet. 

\textbf{Bandwidth.} The bandwidth overhead introduced by the defense is affordable. 
If $\epsilon$ decreases, the latency increases but bandwidth overhead decreases. 
The reason is that when privacy parameter is lower, noise generated by differential privacy is higher, which causes more buffered data per packet and therefore a longer latency on average. On the other hand, more buffered real data are sent by dummy packets generated by adaptive padding, which reduces the overall dummy data needed in each traffic trace.   


\begin{table}[tbp]
\small
	\centering  
	\caption{Tradeoffs with Different Privacy Parameter}
	\begin{tabular}{|c|cc|c|}  
	    \hline
	     Privacy &  \multicolumn{2}{c|}{Latency}  &  
	     Bandwidth
	     \\
	    
	    Parameter $\epsilon$ & Per Packet (ms) & Per Trace (ms) & Overhead (KB) \\
	    \hline 
	    0.005 & 16.5 & 136.0 (2.6\%) & 55.82 (138.7\%)  \\ 
	    \hline 
	    0.05 & 10.4 & 31.4 (0.6\%) & 66.34 (146.0\%)  \\ 
	    \hline 
	    0.5 & 7.0 & 17.4 (0.3\%) & 70.38 (148.8\%)  \\ 
	    \hline
	\end{tabular}
	\vspace{-10pt}
\label{table:defense:overhead}
\end{table}


\section{Limitations and Future Work}
\label{sec:discuss}

\textbf{Human Voices.} In this study, we leveraged automated voices but not human voices to trigger encrypted traffic on a smart speaker during our data collection. One of our future work is to evaluate voice command fingerprinting with human voices by considering different genders, ages and accents.

\textbf{Voice Commands.} 
We did not study popular voice commands that require interactions with other IoT  devices or involve credit card transactions. For instance, asking a smart speaker to successfully order an item online with 1,500 times is challenging to perform in a lab setting. 

We studied 100 popular voice commands in the closed-world setting. On the other hand, we acknowledge that the number of voice commands users could ask in practice is much greater than 100. It would be interesting to assess the privacy leakage of voice command fingerprinting on data with a much greater number of voice commands (e.g., 1,000). A more effective way of collecting data will be needed in that case. 

\textbf{Packet Timing.} We did not leverage packet timing information \cite{RSMGW20} in our attack. 
It would be interesting to examine how packet timing information could be utilized in voice command fingerprinting. We will leave it as a future work.

\textbf{Different Prior Probabilities.}
In this study, as other existing fingerprinting attacks, we assume that each class in the closed-world setting has a uniform prior probability. However, this is not the most accurate way to formulate the problem. Different voice commands could have different prior probabilities if an attacker takes into account additional background information. 

For instance, given two voice commands, \textit{Q1: ``How many days until Thanksgiving?"} and \textit{Q2: ``How many days until Tax Day?"},    
if it is in October, the probability of asking Q1 is obviously greater than the probability of asking Q2. On the contrary, if it is in March, then the probability of asking Q2 is clearly greater than the probability of asking Q1. 
Without the statistic information from voice service providers, accurately formulating the prior probabilities of different voice commands is challenging. 

\section{Conclusion}
\label{sec:conclusion}

We advance the understanding of privacy impacts of smart speakers by investigating voice command fingerprinting attacks using neural networks. Our attack results show worrying privacy concerns especially using incoming traffic only. 
The experimental results show that our proposed defense can mitigate privacy leakage. 
 
\begin{figure*}[ht]
\centering
\includegraphics[width=14cm]{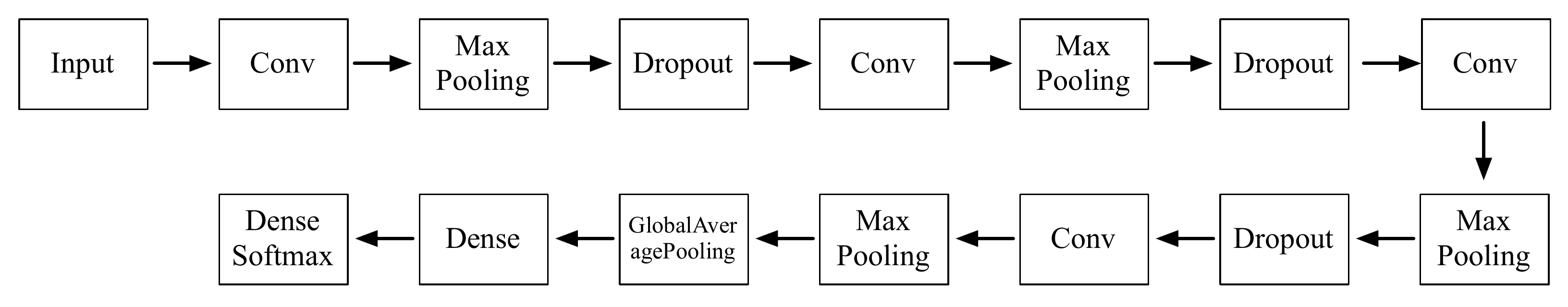}
\vspace{-10pt}
\caption{The structure of our CNN.}
\label{fig:cnn}
\end{figure*}

\begin{figure*}[ht]
\centering
\includegraphics[width=12cm]{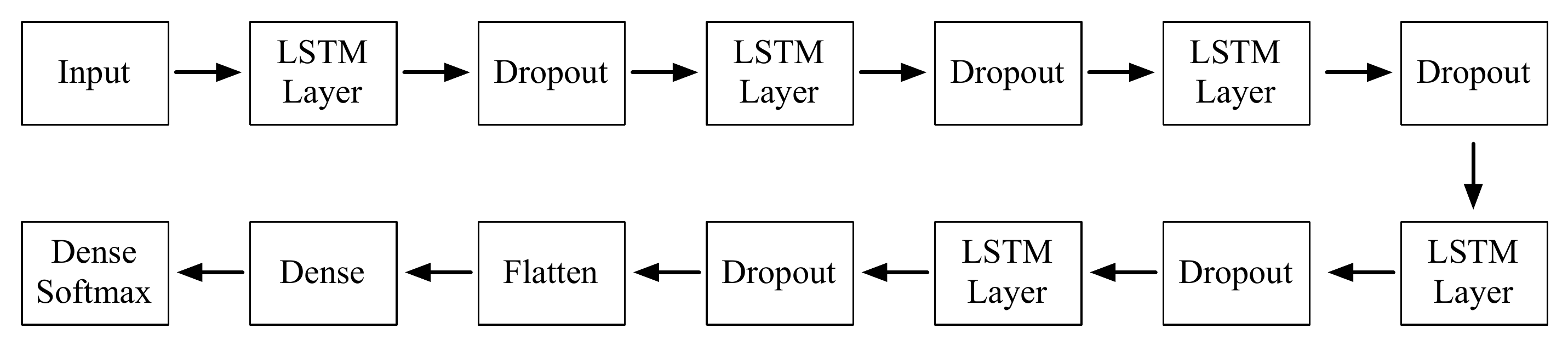}
\vspace{-10pt}
\caption{The structure of our LSTM.}
\vspace{-10pt}
\label{fig:lstm}
\end{figure*}

\begin{table*}[htbp]
\footnotesize
\vspace{5pt}
	\centering  
	\caption{Tuned Hyperparameters of Each Neural Network with Both Outgoing and Incoming Traffic in the Numerical Format}
	\resizebox{\textwidth}{!}{
	\begin{tabular}{|c|c|c|c|c|}  
		\hline
	Hyperparameters  & Search Space & CNN & LSTM &  SAE \\ \hline
	Input Dimension & \{300, 325, 350, ..., 575, 600\}  & 475 & 350  & 375 \\ 
	Optimizer & \{Adam, SGD, Adamax, Adadelta\}  & Adamax & Adamax  & Adam \\ 
	{Learning Rate} & \{0.001, 0.002, 0.01, 0.05, 0.1\}  & 0.002 & 0.002 & 0.001 \\ 
    {Decay} & \{0.00, 0.01, 0.02, ..., 0.50\}  & 0.13 & 0.19  & 0.30 \\ 
	Batch Size & \{30, 40, 50, ..., 120, 130\}  & 70 & 130  & 110 \\ 
	Activation Function & \{softsigh, tanh, elu, selu\}  & [tanh; elu; elu; selu $||$ selu] & [tanh; tanh; tanh; tanh $||$ selu]  & [elu; tanh; selu; elu; softsign $||$ tanh] \\ 
	Dropout & \{0.0, 0.1, 0.2, 0.3, 0.4, 0.5\}  & [0.1; 0.3;  0.1; 0.0] & [0.4; 0.1; 0.1; 0.3; 0.5]  & [0.2; 0.0; 0.0; 0.3] \\
	Dense Layer Size & \{100, 110, 120, ..., 170, 180\}  & 180 & 70  & 130 \\
    Convolution Number & \{16, 32, 64, 128, 256\} & [128; 128; 64; 256] & $-$ & $-$ \\
	Filter Size & \{7, 9, 11, ..., 25, 27\}  & [7; 19; 13; 23] & $-$  & $-$ \\
	Pool Size & \{1, 3, 5, 7\}  & [1; 1; 1; 1] & $-$  & $-$ \\
	LSTM Layer Size & \{90, 100, 110,..., 300, 310\} & $-$ & [210; 190; 190; 190; 130] & $-$ \\
	SAE Encoder Layer Size & \{200, 210, ..., 390, 400\} & $-$ & $-$ & [330; 260; 330; 280; 250] \\
	\hline
	\end{tabular}}
	\vspace{-10pt}
\label{table:attack:hyperparameter:numeric}
\end{table*} 

\begin{table*}[htbp]
\vspace{5pt}
	\centering  
	\caption{Tuned Hyperparameters of Each Model with Incoming Traffic Only in the Numerical Format}
	\resizebox{\textwidth}{!}{
	\begin{tabular}{|c|c|c|c|c|}  
		\hline
	Hyperparameters         & Search Space                          & CNN                                   & LSTM                                  & SAE \\ \hline
	Input Dimension         & \{300, 325, 350, ..., 575, 600\}      & 450                                   & 500                                   & 350 \\ 
	Optimizer               & \{Adam, SGD, Adamax, Adadelta\}       & Adam                                  & Adamax                                & Adadelta \\ 
	{Learning Rate}         & \{0.001, 0.002, 0.01, 0.05, 0.1\}     & 0.002                                 & 0.002                                 & 1.0 \\ 
    {Decay}                 & \{0.00, 0.01, 0.02, ..., 0.50\}       & 0.50                                  & 0.20                                  & 0.30 \\ 
	Batch Size              & \{30, 40, 50, ..., 120, 130\}         & 150                                   & 170                                   & 130 \\ 
	Activation Function     & \{softsigh, tanh, elu, selu\}         & [tanh; selu; elu; selu $||$ selu]     & [tanh; tanh; tanh; tanh $||$ elu]     & [elu; selu; selu; softsign; tanh $||$ elu] \\ 
	Dropout                 & \{0.0, 0.1, 0.2, 0.3, 0.4, 0.5\}      & [0.2; 0.1; 0.4; 0.5]                  & [0.1; 0; 0.1; 0; 0.1, 0.5]            & [0.1; 0.0; 0.0; 0.0] \\
	Dense Layer Size        & \{100, 110, 120, ..., 170, 180\}      & 140                                   & 150                                   & 160 \\
    Convolution Number      & \{16, 32, 64, 128, 256\}              & [256; 32; 128; 32]                    & $-$                                   & $-$ \\
	Filter Size             & \{7, 9, 11, ..., 25, 27\}             & [9; 9; 11; 15]                        & $-$                                   & $-$ \\
	Pool Size               & \{1, 3, 5, 7\}                        & [3; 2; 1; 2]                          & $-$                                   & $-$ \\
	LSTM Layer Size         & \{90, 100, 110,..., 300, 310\}        & $-$                                   & [170; 290; 170; 90; 250]              & $-$ \\
	SAE Encoder Layer Size  & \{200, 210, ..., 390, 400\}           & $-$                                   & $-$                                   & [330; 290; 270; 250; 220] \\
	\hline
	\end{tabular}}
\label{table:attack:hyperparameter:numeric:incoming}
\end{table*} 

\begin{figure*}[ht]
\centering
\includegraphics[width=11cm]{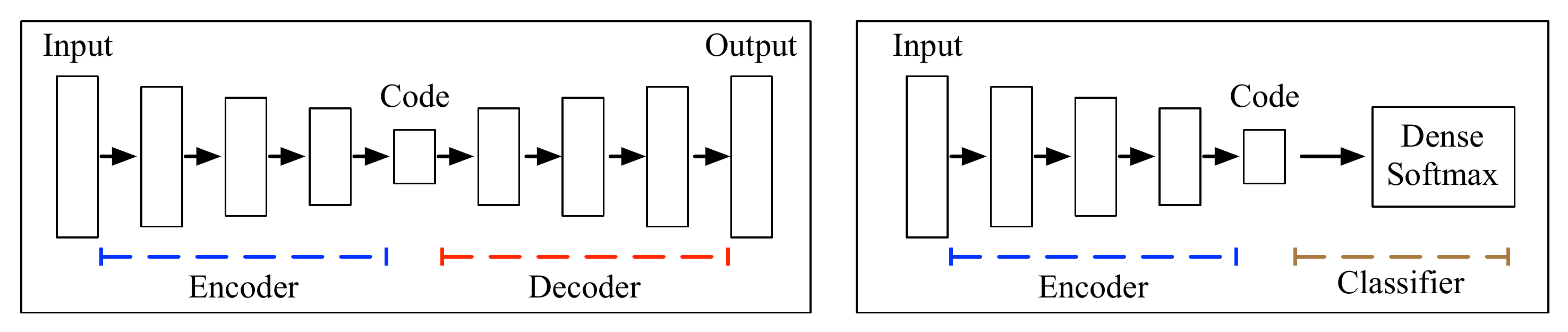}
\vspace{-10pt}
\caption{The structure of our SAE. Encoder and code are trained in training (left) and used in classification (right).}
\label{fig:sae}
\end{figure*}

\section*{Acknowledgements} 
We thank the anonymous reviewers and our shepherd, Dr. Selcuk Uluagac, for their insightful comments on this paper. UC authors were partially supported by National Science Foundation (CNS-1947913), UC Office of the Vice President for Research - Pilot Program, and Ohio Cyber Range at UC. King Hudson was partially supported by NSF LSAMP (Louis Stokes Alliances for Minority Participation) 
program. Wenhai Sun was partially supported by Purdue Research Foundation Summer Faculty Grant. The code, datasets, and other additional information of this study can be found at \cite{DeepVCFingerprinting}. 

\bibliographystyle{ACM-Reference-Format}
\bibliography{main}


\section*{Appendix}
\label{sec:appendix}


\textbf{Tuned Hyperparameters.} 
For the search space of each hyperparameter in Table~\ref{table:attack:hyperparameter:numeric} and Table~\ref{table:attack:hyperparameter:numeric:incoming}, we represent it as a set. We searched for learning rate and decay values if the optimizer is Stochastic Gradient Decent (SGD). If the tuned optimizer is not SGD, we used the default learning rate and decay provided by \texttt{Keras}. For the activation functions, dropout, filter size and pool size, we searched for hyperparameters at each layer. For each of these, the tuned parameters we report in the table are presented as a sequence of values by following the order of layers we presented in Fig.~\ref{fig:cnn}, Fig.~\ref{fig:lstm}, and Fig.~\ref{fig:sae}. For instance, for our CNN, the tuned activation functions are \texttt{tanh} (1st Conv), \texttt{elu} (2nd Conv), \texttt{elu} (3rd Conv) and \texttt{selu} (4th Conv). The \texttt{selu} after symbol $||$ in the table means that the second to last dense layer in our CNN uses \texttt{selu} as its activation function. We did not include \texttt{relu} as one of the activation functions in the search space. It is because \texttt{relu} maps all the negative values (the sizes of all the incoming packets) to 0s, which is not suitable for traffic analysis and has been pointed out in a previous study \cite{SIJW18}.  
Due to space limitation, we skip the tuned hyperparameters in the binary format. 

\begin{table*}[ht]
\footnotesize
\caption{The monitored list of voice commands in the closed-world setting.}
\centering
\begin{tabular}{|c|p{0.4\linewidth}|c|p{0.4\linewidth}|}
\hline
\textbf{Index} & \textbf{Voice Command} & \textbf{Index} & \textbf{Voice Command} \\
\hline
1 & Are you wearing green? & 51 & What is gluten?\\
\hline
2 & Announce Happy Valentines Day. & 52 & What is Homecoming about?\\
\hline
3 & Do dogs dream? & 53 & What is my sports update?\\
\hline
4 & Do you like cats or dogs? & 54 & What is my traffic report?\\
\hline
5 & Flip a coin. & 55 & What is on your mind?\\
\hline
6 & Give me a dinosaur fact. & 56 & What is Roblox?\\
\hline
7 & Give me a fun fact about sleep. & 57 & What is the AFC North Standings?\\
\hline
8 & Good Morning. & 58 & What is the best comedy movie?\\
\hline
9 & Help. & 59 & What is the capital of Spain?\\
\hline
10 & How deep is the Indian Ocean? & 60 & What is the date tomorrow?\\
\hline
11 & How do you spell appreciate? & 61 & What is the fourth book in the Narnia series?\\
\hline
12 & How far away is the moon? & 62 & What is the history of Labor Day?\\
\hline
13 & How hot is the sun? & 63 & What is the longest word?\\
\hline
14 & How many days are in September? & 64 & What is the number one song this week?\\
\hline
15 & How many days in a year? & 65 & What is the price of bitcoin?\\
\hline
16 & How many days until Christmas? & 66 & What is the scariest movie of all time?\\
\hline
17 & How many days until Thanksgiving? & 67 & What is the score of the Eagles game?\\
\hline
18 & How many fantasy points does LeBron James have? & 68 & What is the score of the Red Sox game?\\
\hline
19 & How many ounces in a pound? & 69 & What is the time in Singapore?\\
\hline
20 & How many seconds are in a year? & 70 & What is the weather for Sunday?\\
\hline
21 & How many teaspoons are in a tablespoon? & 71 & What is the weather?\\
\hline
22 & How much does an elephant weigh? & 72 & What is trending?\\
\hline
23 & How much is an ounce of gold? & 73 & What is your favorite flower?\\
\hline
24 & How old are you? & 74 & What is your favorite game?\\
\hline
25 & How old is Henry Winkler? & 75 & What is your favorite hobby?\\
\hline
26 & How old is Serena Williams? & 76 & What is your favorite sport?\\
\hline
27 & How tall is Steph Curry? & 77 & What is your mission?\\
\hline
28 & How tall is the Empire State Building? & 78 & What is zero divided by zero?\\
\hline
29 & How tall is The Rock? & 79 & What movies are playing?\\
\hline
30 & Is a tomato a fruit or a vegetable? & 80 & What were yesterdays scores?\\
\hline
31 & Pick a number? & 81 & When does daylight saving time end?\\
\hline
32 & Surprise me. & 82 & When does Game of Thrones return?\\
\hline
33 & Talk like a pirate. & 83 & When is Boxing Day?\\
\hline
34 & Tell me a barbecue joke. & 84 & When is Hanukkah?\\
\hline
35 & Tell me a coffee joke. & 85 & When is the NBA all star game?\\
\hline
36 & Tell me a fun fact. & 86 & When is the next full moon?\\
\hline
37 & Tell me a Halloween hack. & 87 & Where did Yoda live?\\
\hline
38 & Tell me a joke. & 88 & Where is Mount Rushmore?\\
\hline
39 & Tell me a palindrome & 89 & Who do you love?\\
\hline
40 & Tell me a Star Wars joke. & 90 & Who is in Mastodon?\\
\hline
41 & Tell me some good news. & 91 & Who is nominated for best actor?\\
\hline
42 & Tell me something weird. & 92 & Who is playing Monday Night Football?\\
\hline
43 & Translate good morning to Spanish. & 93 & Who is second in the NBA Western Conference?\\
\hline
44 & What are some flower shops nearby? & 94 & Who is winning the World Series?\\
\hline
45 & What are the most popular books this week? & 95 & Who is your favorite author?\\
\hline
46 & What are the standings in the English Premier League? & 96 & Who is your favorite poet?\\
\hline
47 & What are you thankful for? & 97 & Who is your favorite superhero?\\
\hline
48 & What can you do? & 98 & Who scored for the Golden Knights?\\
\hline
49 & What happened in the midterm elections? & 99 & Why do leaves change color in the fall?\\
\hline
50 & What is brief mode? & 100 & Will it rain tomorrow?\\
\hline
\end{tabular}
\label{table:closed:world:list}
\end{table*}

\begin{table*}[ht]
\footnotesize
\caption{The unmonitored list of voice commands in the open-world setting.}
\centering
\begin{tabular}{|c|p{0.4\linewidth}|c|p{0.4\linewidth}|}
\hline
\textbf{Index} & \textbf{Voice Command} & \textbf{Index} & \textbf{Voice Command} \\
\hline
1 & Are you skynet? & 51 & What are you reading? \\
\hline
2 & Beatbox for me & 52 & What can I do with more than one Echo device? \\
\hline
3 & Can you auto-tune? & 53 & What is 90 degrees Fahrenheit in Celsius? \\
\hline
4 & Can you do an impression? & 54 & What is Carrie Underwood's net worth? \\
\hline
5 & Can you rap? & 55 & What is pi? \\
\hline
6 & Convert one pound to ounces & 56 & What is the dollar to euro exchange rate? \\
\hline
7 & Define lexicon & 57 & What is Don't Worry, He Won't Get Far on Foot movie about? \\
\hline
8 & Do the hokey pokey? & 58 & What is the next book by Rachel Hollis? \\
\hline
9 & Drum roll please & 59 & What is the stock price of General Motors? \\
\hline
10 & Give me a blooper & 60 & What is the tallest animal? \\
\hline
11 & Give me a palindrome & 61 & What is the tallest mountain? \\
\hline
12 & Give me a patriots burn. & 62 & What is the upcoming book by Neil Gaiman? \\
\hline
13 & Give me a prank & 63 & What languages can you translate? \\
\hline
14 & Give me a shark limerick & 64 & What should I be for Halloween? \\
\hline
15 & Give me some bad poetry. & 65 & What time are the Emmys? \\
\hline
16 & How are you? & 66 & What time does L.A. Fitness close? \\
\hline
17 & How did Dow Jones do today? & 67 & What time is Big Brother on TV? \\
\hline
18 & How do I keep my family in sync? & 68 & What is the birthday roundup? \\
\hline
19 & How do I play music everywhere? & 69 & What is the most popular TV show? \\
\hline
20 & How do I say happy birthday in Korean? & 70 & What is the net worth of Jennifer Lawrence? \\
\hline
21 & How do you say happy birthday in Chinese? & 71 & What is your favorite word? \\
\hline
22 & How does the intercom work? & 72 & When do I have to register to vote? \\
\hline
23 & How many calories are in a donut? & 73 & When does Avengers: Endgame release in theaters? \\
\hline
24 & How many gallons of water are in the Atlantic Ocean? & 74 & When does fall begin? \\
\hline
25 & How many hits did Derek Jeter have in 2012? & 75 & When does the new season of The Walking Dead premiere? \\
\hline
26 & How many people live in China? & 76 & When does Wimbledon start? \\
\hline
27 & How many people live in New York?  & 77 & When is April Fools Day? \\
\hline
28 & How many rushing yards did Emmitt Smith have in his career? & 78 & When is Shark Week? \\
\hline
29 & How many times has Duke been to the Final Four? & 79 & When is the autumn equinox? \\
\hline
30 & How much does a Lamborghini cost? & 80 & When is the next lunar eclipse? \\
\hline
31 & How old is Queen Elizabeth? & 81 & When is the Patriots' first game? \\
\hline
32 & Is a hot dog a sandwich? & 82 & When is the Tour de France? \\
\hline
33 & Make animal noises & 83 & When was McDonald's founded? \\
\hline
34 & Pretend to be a supervillain & 84 & Where is the MLB All-Star game being played? \\
\hline
35 & Rap about the cloud & 85 & Who does Alabama play in their first game? \\
\hline
36 & Recite a haiku & 86 & Who inspires you? \\
\hline
37 & Release the Kraken & 87 & Who invented GPS? \\
\hline
38 & Roll the dice & 88 & Who is leading the Players Championship? \\
\hline
39 & Sing a song for the 4th of July & 89 & Who is Peppa Pig? \\
\hline
40 & Star Wars or Star Trek & 90 & Who is Rainbow Dash? \\
\hline
41 & Tell me a baseball story & 91 & Who is running for Senate in California?\\
\hline
42 & Tell me a basketball joke & 92 & Who is your favorite baseball player? \\
\hline
43 & Tell me a giraffe fact & 93 & Who is your favorite Pokémon? \\
\hline
44 & Tell me a joke about you & 94 & Who leads the WNBA in scoring? \\
\hline
45 & Tell me a pun & 95 & Who signed the Declaration of Independence? \\
\hline
46 & Tell me a shark joke & 96 & Who stars in Gringo? \\
\hline
47 & Tell me a tongue twister & 97 & Who won the Winter Classic? \\
\hline
48 & What are some good Halloween movies to watch? & 98 & Who is going to win the Final Four? \\
\hline
49 & What are the best books of the year so far? & 99 & Who is hosting Saturday Night Live this weekend? \\
\hline
50 & What are the rarest skins in Fortnite? & 100 & Who is your favorite college basketball team?\\
\hline
\end{tabular}
\label{table:open:world:list}
\end{table*}

\begin{figure*}[t]
  \centering
  \begin{subfigure}[b]{0.48\textwidth}
    	 \centering
         \includegraphics[width=\textwidth]{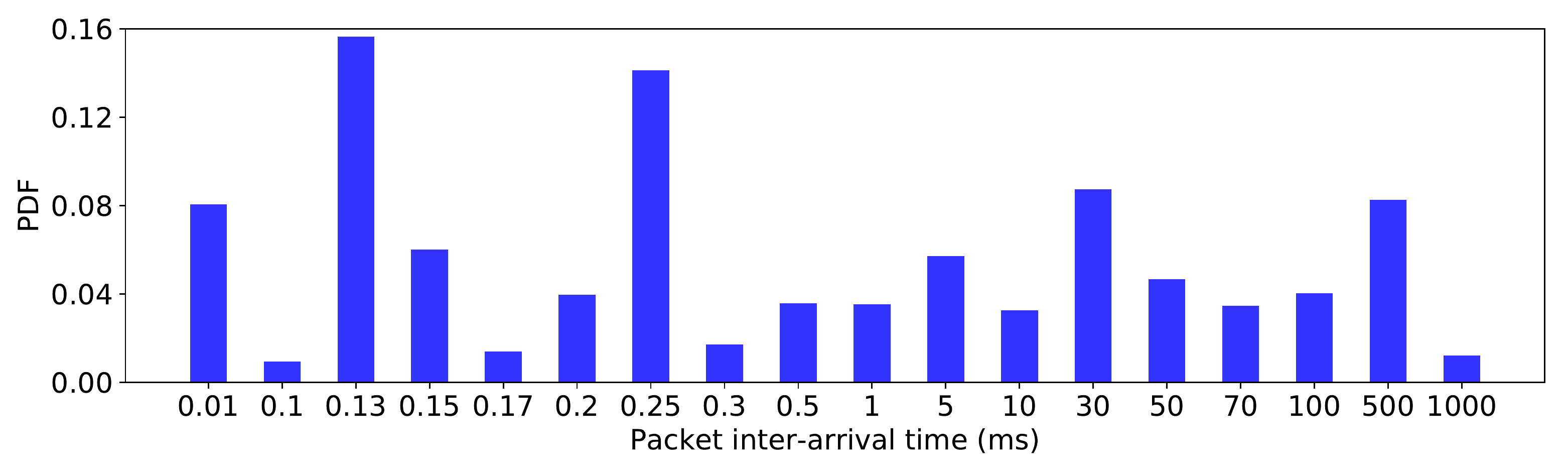}
         \caption{Inter-arrival time distribution of incoming packets.}
     \end{subfigure}
     \hfill
     \begin{subfigure}[b]{0.48\textwidth}
        \centering
        \includegraphics[width=\textwidth]{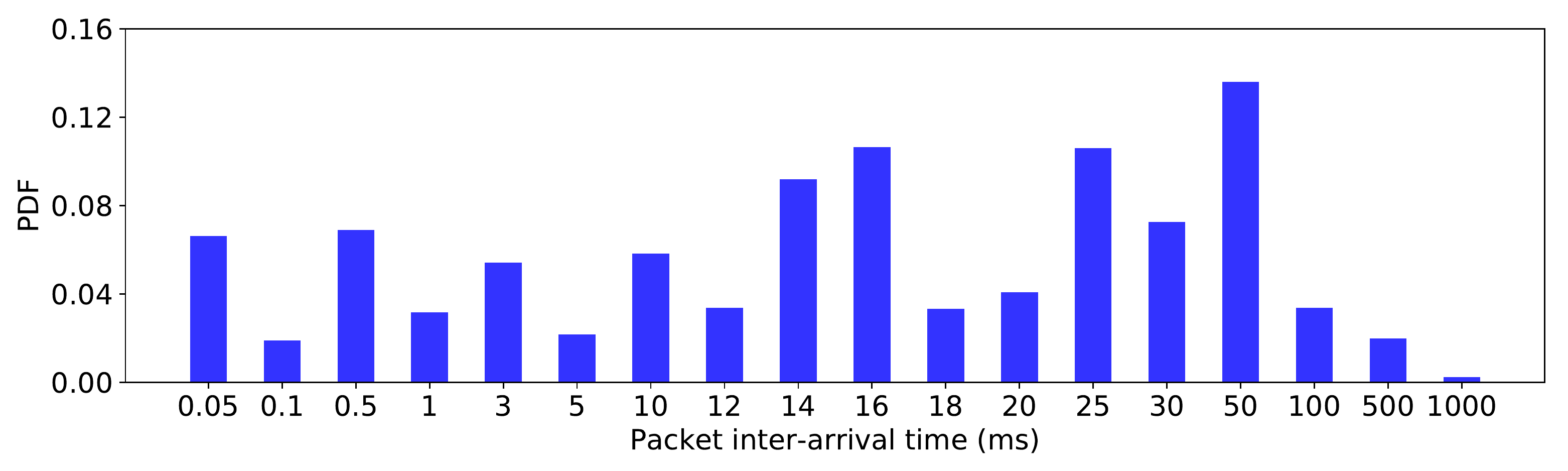}
        \caption{Inter-arrival time distribution of outgoing packets.}
     \end{subfigure}
     \caption{The distribution of packet inter-arrival time.}
     \label{fig:packet:interval}
     \vspace{-10pt}
\end{figure*}  

\textbf{Packet Inter-arrival Time.} Fig.~\ref{fig:packet:interval} describes the distribution of packet interarrival time in Amazon Echo dataset. We leveraged this distribution to perform adaptive padding in our defense.

\end{document}